\documentclass[prd,aps,twocolumn,floatfix,10pt]{revtex4-1}
\usepackage{lipsum}
\usepackage{textcomp}
\usepackage{amsmath}
\usepackage{amsfonts}
\usepackage{amssymb}
\usepackage{graphicx}
\usepackage{hyperref}
\usepackage{subfigure}
\usepackage[usenames]{color}

\begin{document}

\title{Multi-solitons with vector mesons on the two-sphere}

\author{F. L. Carrasco}
\email{fedecarrasco@gmail.com}
\author{O. A. Reula}
\email{oreula@gmail.com}
%\homepage{http://legauss.blogspot.com}
\affiliation{ FaMAF-UNC, IFEG-CONICET, Ciudad Universitaria, 5000, C\'o{}rdoba, Argentina.}

%\author{W. Y. S. I. W. Y. Get}
%\email{wisiwig@duh.com}
%\homepage{http://www.google.com}
%\affiliation{National Other Institute of Other Country.}

\date{\today}

\begin{abstract}

Recent studies have suggested a strong connection between the static solutions of the 3D Skyrme model and those corresponding to its low-dimensional analog (baby-Skyrme 
model) on a two-sphere. We have found almost identical solutions considering an alternative two-dimensional model in which a vector meson field
is introduced and coupled to the system, instead of the usual Skyrme term.  
It has been known that including this vector meson field in three dimensions stabilizes the non-linear sigma model without the need of a 
term quartic on derivatives of the pion fields (Skyrme term).
In the present work, we have numerically searched for static multi-solitonic solutions of this alternative stabilization, for the case in which the base-space is a two-sphere.
Moreover, we analyze the stability of these solutions under small perturbations in a fully dynamical setting.

We have also considered the inclusion of a particular potential term into the Lagrangian, and explored the low/high-density phases of solitons for different ranges of the 
parameter space, achieving solitons localized enough that allow us to compare with planar (two-dimensional) studies.
\end{abstract}

%\keywords{}

\maketitle

\section{Introduction}

A sigma model is a non-linear field theory where the fields takes values on a Riemannian manifold. 
That is, a map from a space-time into a target space, usually of higher symmetry. 
It is one of the simplest systems admitting static topological soliton solutions, which can be characterized
by the degree of the map: an integer $B$ that in the field theory 'language' is known as the \textit{topological charge} of the field configuration. 
Identifying this scalar field (or map) with a pion field in three spatial dimensions, the different topological solitonic solutions can be interpreted as baryons.\\
However, the topologically non-trivial static solutions are dynamicaly unstable, and the model requires the inclusion of 
extra terms into the lagrangian in order to yield stable solutions.
With only pion degrees of freedom, the traditional inclusion to the lagrangean is given by the so called Skyrme term, which is quartic in derivatives. 
This leads to the well known Skyrme model \cite{ Skyrme}.\\
Within the Skyrme model, a soliton with topological charge one is called a \textit{Skyrmion}, 
which suitably quantized constitutes a model for a physical nucleon.
While solitons of higher topological charges (\textit{multi-Skyrmions}), are classical models for higher nuclei. 
This idea of representing nucleons as solitons of the effective pion field remains attractive even in the context of QCD, and moreover,
the approach becomes exact in the large $N_c$ limit \cite{ Witten}.

It was later realized, that it is possible to stabilize the non-linear sigma model without this
fourth order term, by coupling the baryon current to a $\omega$ meson field \cite{ Nappi}.
This new model has a few important advantages over the original one. The Skyrme model, being a quasi-linear system of equations, 
has propagation speeds which depend on the solution,and they can, for some initial data, become imaginary,
destroying the well posedness of its evolution and so the predictive power of the theory. 
While, in the other hand, the vector meson model is a semi-linear, symmetric hyperbolic system of equations. 
And thus, it has all propagation speeds proportional to the speed of light.
Later extensions of this new proposal to stabilize the sigma model, have considered multisolitons in three spatial dimensions \cite{ Sutcliffe-3d} 
and in its two-dimensional analog \cite{ Sutcliffe-2d}. 
There were also a couple of studies on the dynamical aspects of the model \cite{ Amado-1, Amado-2}. 
All of these works have shown that the vector meson and the Skyrme models have very similar properties.

Usually, topological solitons are studied on flat space, but there are various reasons why the curved-space setting is interesting.
The most important one being the emergence of a new length scale on the problem, namely, the one of the underlying geometry. 
Allowing for an interesting interplay between this new scale and the size of the solitons. 
Physically, it allows to model a finite density of solitons, and the transition between the high-density and the low-density phases, 
which traditionally would have involve the rather cumbersome numerical task of putting the skyrmions on a lattice. 
It was realized that many of the qualitative results of lattice calculations could be
obtained in a much easier way by studying the behavior of few solitons on a compact manifold \cite{ Manton-2, Manton-3}. 
This feature has shown to have several potential applications in condensed matter physics (e.g. ref. \cite{Sondhi, Belavin}).

In this work, we want to study this alternative stabilization of the sigma model by means of the inclusion of a $\omega$ vector meson, 
in its two-dimensional version on a unit sphere.
In ref. \cite{ Hen}, the authors have studied the baby-Skyrme model on the two-sphere, and a strong connection between the symmetries of their solutions 
(on the sphere) and those of the 3D Skyrme model was pointed out.
We want to numerically find the static multi-soliton solutions of our model and compare them with the ones found on \cite{ Hen} for the baby-skyrme 
model and with those obtained in \cite{ Sutcliffe-3d} for the 3D vector meson theory. Also, we want to further explore the role of the ratio 
(size of the solitons)/(radius of the sphere) on the possible static configurations of the model.

We begin by describing the model in some detail on Section II. 

Section III discusses the initial configurations considered, as well as details of our numerical implementation.

In Section IV we show the results of our computational studies. First, we treat the case in which no potential term 
is present, like the one in Ref. \cite{ Hen} for the baby-skyrme model. 
Then, we include a potential and consider the interplay of the two length scales now presents. 

We summarize and conclude with some further comments on Section V.

\section{Formalism}

\subsection{The Model}

Our starting point is the non-linear sigma model from $S^2$ to $S^2$. That is, a map $\phi$ between a space-time ($\mathbb{R}\times S^2$, $g$)
and a target Riemannian space ($S^2$, $H$), where $g_{ab}= h_{ab} - n_{a}n_{b}$ ($n_{a}$ being the normal to the $t=const$ homogeneity hypersurfaces) and 
with both $h_{ab}$ and $H_{AB}$ representing the metric of the unit sphere.
We will use capital letters to denote indices on the target space and the lower case to represent  spacetime indices.

The action is given by the integral (over a space-like hypersurface) of the trace of the pull back of the Riemannian metric (target space) 
into the space-time (base space):
\begin{equation} 
 S(\phi):= \frac{1}{2} \int_{\Sigma_{t}} g^{ab} \nabla_{a}  \phi^{A}  \nabla_{b}  \phi^{B} H_{AB}  dV 
\end{equation}

The topological degree of the map is an integer $B$ that characterizes the field configuration at a given time.
It is defined as the pull back of the surface element of the target $S^2$, integrated over the physical space and normalized by the total area,
\begin{equation} 
 B = deg[\phi] := \frac{1}{4\pi} \int_{\Sigma_{t}} \Omega_{ab}
\end{equation}
Where $\Omega_{ab}:= \frac{1}{2} \nabla_{a} \phi^{A} \nabla_{b} \phi^{B} \varepsilon_{AB} $ is, as mentioned, the pull-back 
of the surface element $\epsilon_{AB}$ of the target manifold.

Since the degree of the map $B$ is integer-valued quantity, it turns out that 
it must be conserved throughout the evolution, assuming the dynamics to be smooth (i.e: no singularities developed).
This also follows directly from the conservation of the topological current $B^{a}$ (i.e: $\nabla_a B^a = 0$), which is just 
 the \textit{hodge dual} of the two-form $\Omega_{ab}$,
\begin{equation} 
 B^{a} := -\frac{1}{4\pi}\epsilon^{abc}\Omega_{bc} %= -\frac{1}{8\pi} \epsilon^{abc} \epsilon^{ABC} \phi_A (\nabla_{b}\phi_B)  (\nabla_{c} \phi_C)
\end{equation}

It might be convenient to think the map $\phi$ as taking values on $\mathbb{R}^3$, but constrained to the sphere. 
That is, $\phi_{A}$ where the index $A$ takes values $A=1,2,3$ and with $\phi^A \phi_A := \phi_{1}^2 + \phi_{2}^2 + \phi_{3}^2 = 1$. \\  
In this constrained formulation, the Lagrangian density reads,
\begin{equation} 
 \mathcal{L}_{\sigma} = \frac{1}{2} g^{ab} (\nabla_{a} \phi^A ) (\nabla_{b} \phi^B ) \delta_{AB} + \frac{1}{2}\lambda(1- \phi^{2})
\end{equation}
where $\lambda $ is the Lagrange multiplier enforcing the constraint. 

Following references \cite{ Nappi, Sutcliffe-2d}, and in order to stabilize the solitonic solutions of this theory, we introduce a vector meson field 
$\omega$ coupled through the baryon current, and obtain the Lagrangian density of the model:  
\begin{equation} 
 \mathcal{L} := \mathcal{L}_{\sigma} + V(\vec{\phi}) + \frac{1}{2} \nabla_{a}\omega_{b}(\nabla^{a}\omega^{b} - \nabla^{b}\omega^{a}) + \frac{1}{2}M^{2}\omega_{a}\omega^{a} + g\omega_{a} B^{a}
\label{lagrangean}
\end{equation}
where $M$ is the mass of the meson field and $g$ is the coupling constant. We have also included a potential term $V$ which only depends on the pion field, $\phi^{A}$. 

\subsection{Equations of Motion}

Variations of the action with respect to the scalar and vector fields leads to the usual Euler-Lagrange equations. When applied to the 
Lagrangian \eqref{lagrangean}, it gives
\begin{equation}
\Box\omega_{a} - \nabla^{b}\nabla_{a}\omega_{b} - M^{2}\omega_{a} - gB_{a} = 0  
\label{ec-w}
\end{equation}
\begin{equation} 
 \Box\phi^{A} + \lambda\phi^{A} + \frac{g}{8\pi}\epsilon^{abc}\epsilon^{ABC} \left[ 3\omega_{a}\nabla_{b}\phi_{B} + 2 \phi_{B} \nabla_{b}\omega_{a} \right]\nabla_{c}\phi_{C} = \frac{\delta V}{\delta \phi^A}
\label{ec-phi}
\end{equation}
Solving for the Lagrangian multiplier we get,
\begin{equation} 
  \lambda = \nabla^{a} \phi_{A} \nabla_{a} \phi^{A} + 3g\omega_{a} B^{a} + \frac{\delta V}{\delta \phi^A} \phi^A
\label{lamba-w}
\end{equation}
which plugged on \eqref{ec-phi} leads to,
\begin{widetext}
\begin{equation}
 \phi^{A}_{tt} = H^{A}_{D} \left\lbrace \Delta \phi^{D} + \frac{g}{8\pi} \epsilon^{abc}\epsilon^{DBC} ( 3\omega_{a}\nabla_{b}\phi_{B} + \phi_{B} F_{ba} ) \nabla_{c}\phi_{C} + \left( \frac{\delta V}{\delta \phi^B} \phi^B \phi^D - \frac{\delta V}{\delta \phi_D}\right)  \right\rbrace  - \phi^{A}\phi^{D}_{t}\phi_{Dt}\nonumber
\end{equation}
\end{widetext}
where $ H^{A}_{B} = \delta^{A}_{B} - \frac{\phi^{A} \phi_{B}}{\phi^{2}} $ is the projector on the plane perpendicular to $\phi^{A}$. \\
Note that the term $ \frac{3g}{8\pi} \epsilon^{abc}\epsilon^{ABC}\omega_{a}\nabla_{b}\phi_{B} \nabla_{c}\phi_{C} $ is parallel to $\phi^{A}$, so
its projection will be zero and can be removed from the last equation. 

From the remaining equation of motion \eqref{ec-w} one can see that the mass $M$ breaks the gauge symmetry, 
enforcing the Lorentz gauge (i.e: $\nabla^{a} \omega_{a} = 0$). Thus, the principal part of the $\omega_a$ equation is just the wave equation.

\subsection{Evolution Equations}

In order to evolve the system, we bring it to a system of first order evolution equations. We choose a set of dynamical variables 
that we believe are the most convenient for doing this: 
for the scalar fields, we chose the field themselves and their first derivatives. 
That is, $\phi^{A}$ and $ \phi^{A}_{a} \equiv \partial_{a} \phi^{A}$.
While for the vector field, we consider $\omega_{a}$, $ F:= \frac{1}{2} \epsilon^{ij} F_{ij} $ and 
$ E_{i}:= F_{i0} $ (where $ F_{ab}:= \nabla_{[a}\omega_{b]} $ is a low-dimensional analog of the electromagnetic tensor).
We close the system by using the integrability condition for $F_{ab}$, namely, $\nabla_{[a}F_{bc]}=0 $.

Thus, we obtain the following set of evolution equations: 
\begin{widetext}
\begin{eqnarray}
\partial_t \phi^{A}_{t} & = & H^{A}_{D} \left\lbrace  \frac{1}{\sqrt{-g}} \partial_{k} ( \sqrt{-g} \phi_{k}^{D}) + \frac{g}{8\pi} \epsilon^{abc}\epsilon^{D}_{\text{  }BC} \phi^{B} F_{ba} \phi_{c}^{C} + \left( \frac{\delta V}{\delta \phi^B} \phi^B \phi^D - \frac{\delta V}{\delta \phi_D}\right)  \right\rbrace  - \phi^{A}\phi^{B}_{t}\phi_{t B} \nonumber\\
\partial_t \phi^{A}_{i} & = & \partial_i \phi^{A}_{t} \nonumber\\
\partial_t E_{i} & = & g^{jk} \nabla_{k} (\epsilon_{ij} F) + M^{2}\omega_{i} + g B_{i} \nonumber\\
\partial_t \omega_{0} & = & g^{ij} \nabla_{i} \omega_{j}  \label{eqs-evol} \\
\partial_t F & = & \epsilon^{ij} \partial_{j} E_{i} \nonumber\\
\partial_t \omega_{i} & = &  \partial_{i}\omega_{0} - E_{i} \nonumber
\end{eqnarray}
\end{widetext}
Subject to the two constraints,
\begin{eqnarray}
g^{jk}\nabla_k E_{j} - M^2 \omega_{0} - g B_{0} & = & 0
\label{vinculo-w}\\
F - \epsilon^{ij} \partial_{i} \omega_{j} & = & 0
\end{eqnarray}

It follows that the set of equations \eqref{eqs-evol} constitute a \textit{strongly hyperbolic system} \cite{ Kreiss}, 
and thus, the Cauchy problem is \textit{well-posed}.

The energy density is found to be,
\begin{widetext}
\begin{equation}
 T_{00} =  \frac{1}{2}\left[ \dot{\phi^A}\dot{\phi_A} + \nabla_k \phi^A \nabla^k \phi_A + 2V(\vec{\phi}) + E_k E^k + F^2 + M^{2} (\omega_{0}^2 + \omega_{k}\omega^{k}) \right] 
\label{energy}
\end{equation}
\end{widetext}
Is important to note here that, although the interaction term doesn't appear explicitly 
in \eqref{energy}, it comes through the constraint \eqref{vinculo-w}  when minimizing the energy functional.

\subsection{Static Solutions}

As mentioned, part of the work is to find the static solutions of the model presented so far: the $\sigma$-model stabilized through the inclusion of a vector field.\\
In a static configuration the spatial components of the baryonic current are zero, because they involve temporal derivatives of the $\phi$ fields.
Thus, given that the current $ B^{i} $ acts as a source for $ \omega_{i} $, one have that the vector field $\omega_i$ should be constant,
and in fact must be null, since it lives on the sphere. 
So $\omega_{i}=0$, and from now on, we will write $ \omega_{0} \equiv \omega $ for notational convenience. 
In a static situation one also have from the definitions that, $ F=0 $ and $ E_k = \partial_k \omega $. 

Performing the above substitutions on equations \eqref{eqs-evol}-\eqref{vinculo-w}, we get the following system of elliptic equations:
\begin{widetext}
\begin{eqnarray}
H^{A}_{D} \left\lbrace \Delta\phi^{D} + \frac{g}{4\pi} \epsilon^{DBC} \epsilon^{ij} \phi_{B}(\partial_{i}\omega) (\partial_{j}\phi_{C}) + \left( \frac{\delta V}{\delta \phi^B} \phi^B \phi^D - \frac{\delta V}{\delta \phi_{D}} \right)  \right\rbrace  & = & 0
\label{est-phi} \\
\Delta\omega - M^2 \omega - g B_{0} & = & 0 
\label{est-w} 
\end{eqnarray}
\end{widetext}

To solve them and get the static solutions, we evolve appropriate initial data along a parabolic flow given by,
\begin{widetext}
\begin{eqnarray}
\partial_{t} \phi^{A}  & = & H^{A}_{D} \left\lbrace \Delta\phi^{D} + \frac{g}{4\pi} \epsilon^{DBC} \epsilon^{ij} \phi_{B}(\partial_{i}\omega) (\partial_{j}\phi_{C}) 
+ \left( \frac{\delta V}{\delta \phi^B} \phi^B \phi^D - \frac{\delta V}{\delta \phi_{D}} \right)  \right\rbrace \label{difusion} \\
\partial_{t} \omega & = & \Delta \omega - M^2 \omega - g B_{0}  \label{difusion-w}
\end{eqnarray}
\end{widetext}

These are essentially heat-type equations (parabolic system), but with some extra terms. We note that if 
the evolution of such system dissipates energy and reach a stationary state, then the R.H.S of \eqref{difusion}-\eqref{difusion-w} would 
be zero and we would be in presence of a solution of \eqref{est-phi}-\eqref{est-w}, namely, a static solution of the theory.   

\section{Computation}

In this study we have used two different codes. One implements the full dynamics of the system through the computation of
the hyperbolic set of equations \eqref{eqs-evol}, while the other serves to explore the possibles static configurations by 
evolving the parabolic system \eqref{difusion}-\eqref{difusion-w}. 
The key feature of having this two codes relies on the possibility of using the output of one of them as initial data for the other. 
Allowing us to check, in a fully dynamical setting (hyperbolic code), if a given configuration found 
after the diffusion process (parabolic code) is really a static and stable solution of the theory.

We have backed up all our numerical results by monitoring conserved quantities like the energy and the topological charge.

\subsection{Grid Scheme}

\begin{figure}[h!]
  \begin{center}
\includegraphics[scale=0.2]{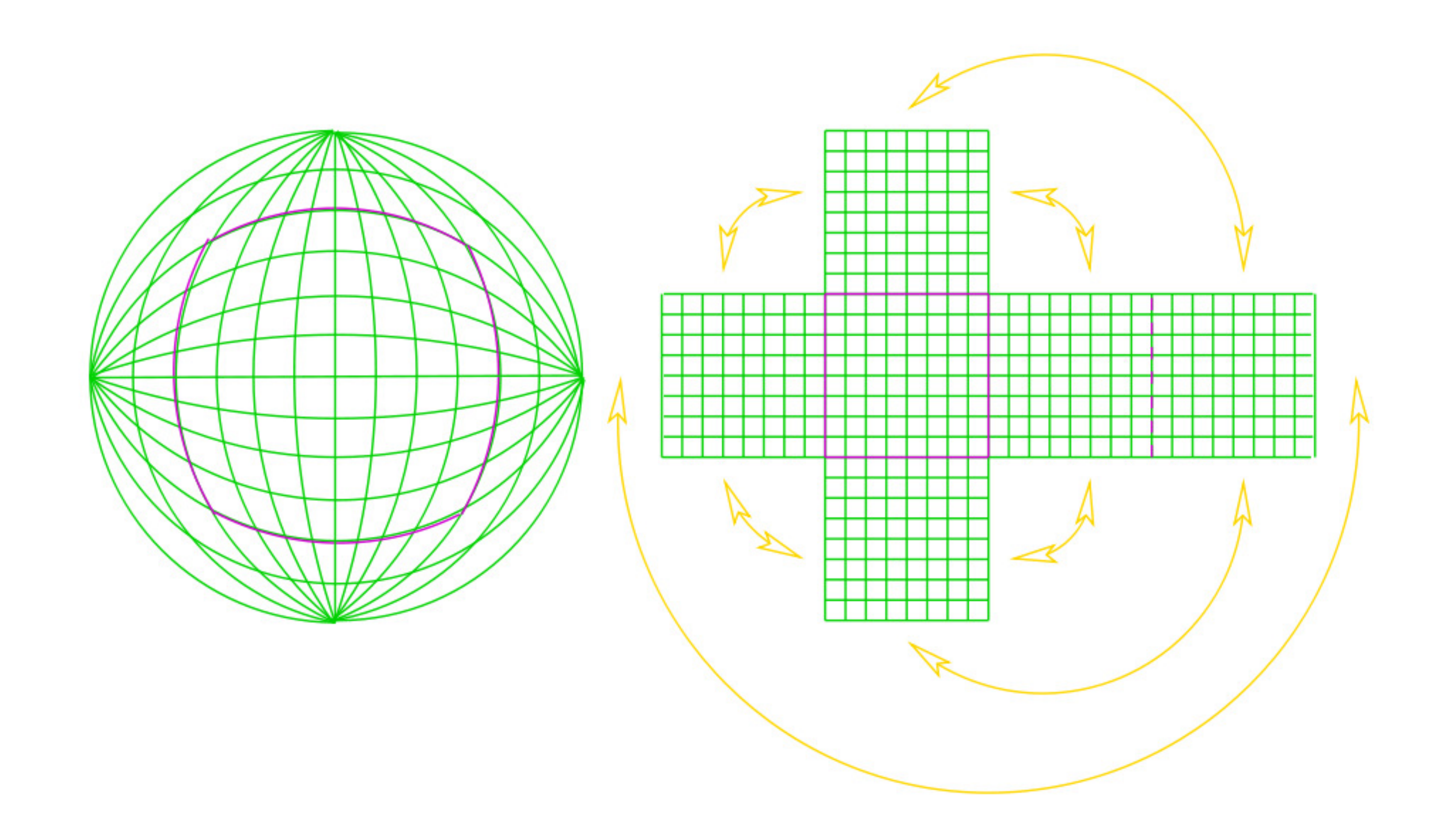}
  \caption{\textit{Cubed Sphere Coordinates.} A total of six Cartesian patches
are employed to cover the sphere. Only patch boundaries coincide at common points.}
 \label{sphere} 
 \end{center}
\end{figure}

As mentioned, the topology of our computational domain is $S^2$, the unit sphere. Since it is not possible to cover the whole sphere 
with a single system of coordinates which is regular everywhere, we employ multiple patches to cover it. \\
A convenient set of patches is defined by the \textit{cubed sphere coordinates}. There are six patches with coordinates 
projected from the sphere, and each of this patches constitute a uniform grid, with a variable amounts of points. (See Fig. \ref{sphere})\\
This grids are defined in a way such that there is no overlap and only grid points at boundaries are common to different grids 
(\textit{multi-block approach}).

Is evident that to solve a problem under this grid structure one must ensure the suitable transfer of information among the different grids. 
We basically follow the technique described on \cite{ Leco}, which relies on the addition of
suitable \textit{penalty terms} to the evolution equations. These terms penalize the possible mismatches between the different values the 
characteristic fields take at the interfaces. For the parabolic case, they were obtain from an extension to the 2-dimensional case by Parisi-Reula (work in progress)
of the studies of Carpenter et al. \cite{ Carpenter1999, Carpenter2001}. 

\subsection{Numerical Scheme and Stability}

In order to construct stable finite difference schemes for our initial value problems we use the method of lines \cite{ Kreiss}.
This means that we first discretize the spatial derivatives (constructing some finite difference operators) so as to obtain a large system of
ordinary differential equations for the grid functions. This system is usually called \textit{semi-discrete system}.

To ensure the stability of the numerical scheme we use the \textit{energy method} described on \cite{ Reula}. First, one have to check that 
the initial value problems are \textit{well-posed} at the continuous level, and that the solutions of the partial differential equations
satisfy an \textit{energy estimate} which bounds some norm of the solution. Then, one construct a difference operator that satisfies 
\textit{summation by parts} (the discrete analogue of integration by parts) in the discrete version of that norm. This operators, together
with the appropriate penalty terms at the interfaces, implies a semi-discrete energy estimate which ensures that the system is stable.
Finally, by discretizing the time derivatives one obtains the fully discrete system which is numerically implemented.
If the semi-discrete system is stable one can show that the fully discrete system is stable as well, provided an appropriate time integrator 
is chosen. We have used for both codes a classical $4^{th}$ order Runge-Kutta.

\subsection{Initial Data}

The initial data was constructed from rational maps, which are the well known static solutions of the sigma model (solutions of the
Bogomolny equations). (For a review on the subject see \cite{ Manton}).\\ 
Rational maps of each topological sector were used to set the initial values for the scalar field.
Written in terms of the target space coordinate $ R = R(z,\bar{z}) $, where $ R = (\phi_1 + i \phi_2)/(1 + \phi_3) $ (and with 
$ z = \tan{\frac{\theta}{2}} e^{i\varphi} $ given in standard angular coordinates), the map used was $ R(z)= \frac{\lambda}{z^B} $,
 with $\lambda \in \mathbb{R}$ and $B$ being the topological degree of the map.
These configurations represents rings of baryon density for all values of charge $B \geq 2$.\\ 
For the vector meson field, we approximate its initial values with $\omega = \frac{g}{M^2} B^0$, that comes from neglecting the Laplacian 
on equation \eqref{est-w}.

As mentioned, for the code that run the full dynamics (hyperbolic), we use the output configurations attained with 
the previous diffusion process (parabolic). Setting the initial values of the remaining dynamical variables (i.e: the absents fields in
the parabolic eq's) to their static values. That is, $\partial_{t} \phi^{A}=0$, $F=0$ and $ E_k = \partial_k \omega $.
%\textbf{No pusimos nada de scattering, estaba pensando si hay algo que compare la dinámica de estos sistemas con la de los skyrmiones en el caso de scattering, o solo hay cosas en lo referente a las configuraciones estáticas?}

%%%%%%%%%%%%%%%%%%%%%%%%%%%%%%%%%%%%%%%%%%%%%%%%%%%%%%%%%%%%%%%%%%%%%%%%%%%%%%%%%%%%%%%%%%%%%%%%%%%%%%%%%%%%%%%%%%%%%%%%%%%%%%%%%%%%%%

\section{Results}

\subsection{$V=0$}

While in flat two dimensional space a potential term $V(\vec{\phi})$ is mandatory to ensure the existence of stable solutions, in the present
model it is not. 
The need for the inclusion of this potential comes from the necessity of avoiding Derrick's scaling argument, 
and it basically determines, together with the other terms added to the Lagrangian of the sigma model, the appearance of a preferred size for the 
solitons. In the case of the unit sphere, however, there already exists a natural spatial scale: the radius of curvature. 
Therefore Derrick's argument doesn't apply here and stable static solutions are in principle allowed.

Thus, it is expected that the solitons, in the absence of a potential term, will be spread out over the entire space.
And in particular, the $B=1$ solution is expected to be the uniform energy (or topological) density distribution on the sphere.
This is in fact the case, and correspond simply to the rational map solution $ R(z)= \frac{1}{z} $, which is
found to satisfy the static equations \eqref{est-phi}-\eqref{est-w}, leading to the configuration $ B^0 = \frac{1}{4\pi} = cte $.
We were able to confirm numerically that this is the static configuration of the sector $B=1$, and to verify that the energy found agrees with 
its theoretical value given by $ E = 4\pi + \frac{1}{4\pi} \frac{g^2}{M^2} $.

The sector $B=1$ seems to be the only topological sector in which the full O(3) symmetry of the theory is preserved.
The $B=2$ solution turned out to be axially symmetric (corresponding to the O(2) subgroup), whereas higher charge multi-solitons were all found 
to have point symmetries which are subgroups of O(3). In order to ensure these results are not influenced by the initial data chosen 
(on the first two sectors) or by the grid structure employed (on the higher charged cases), we have included an artificial numerical perturbation on top 
of the relaxation scheme (parabolic code).
This mechanism consist in the addition, to every fields at every grid point, and for each time step, of a random number chosen from the range $(-1,1)$ 
multiplied by a coefficient $\delta$ (generally chose to be $\delta= 10^{-3}$).
Such mechanism has shown the robustness of the axial symmetry in the $B=2$ sector and give us more confidence that the configurations found are 
not influenced by the numerical implementation. In addition, we have observed that it seems to stimulate the diffusion process as well.
%\textbf{esto puede ser consecuencia de la manera con que dimos los datos iniciales}

In Ref.\cite{ Hen} was pointed out an intimate relation between the baby-skyrme model on the sphere and the 3D Skyrme model. The symmetries
of the 3D Skyrme model are determined solely by the angular dependence of the skyrme field, and it was suggested that this 2D version may
be thought of as the 3D Skyrme model with a 'frozen' radial coordinate. They were able to demonstrate that connection within the rational map 
approximation, and show full field simulations for charges up to $B=14$ having the same symmetries as corresponding solutions of the 3D 
Skyrme model.

\begin{figure*}
\    \ \vfill
\    \ \vfill
%\centering{
\begin{minipage}{5cm}
 \subfigure[ $B=2$]{\includegraphics[scale=0.25]{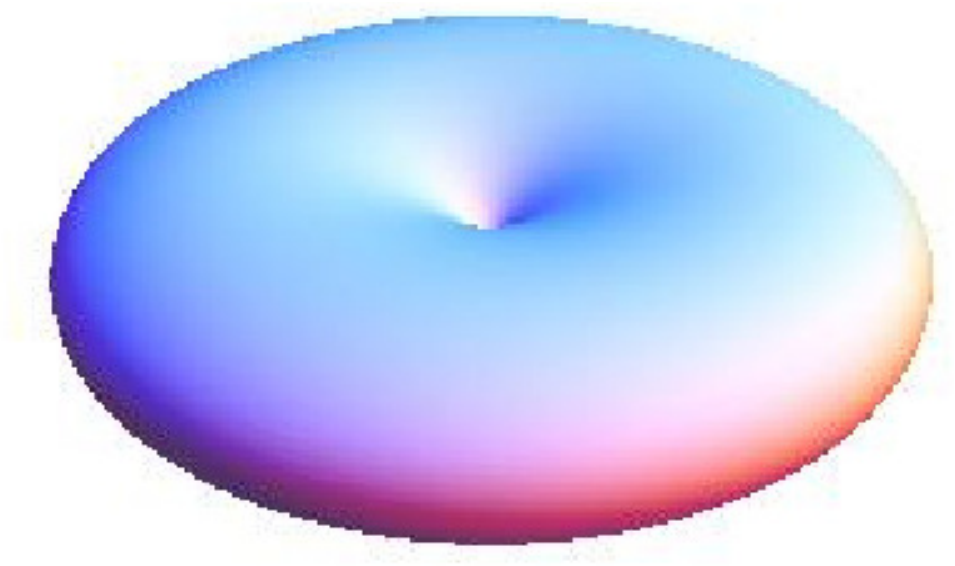}}
\end{minipage}
%\    \ \hfill
\begin{minipage}{5cm}
\subfigure[ $B=3$]{\includegraphics[scale=0.25]{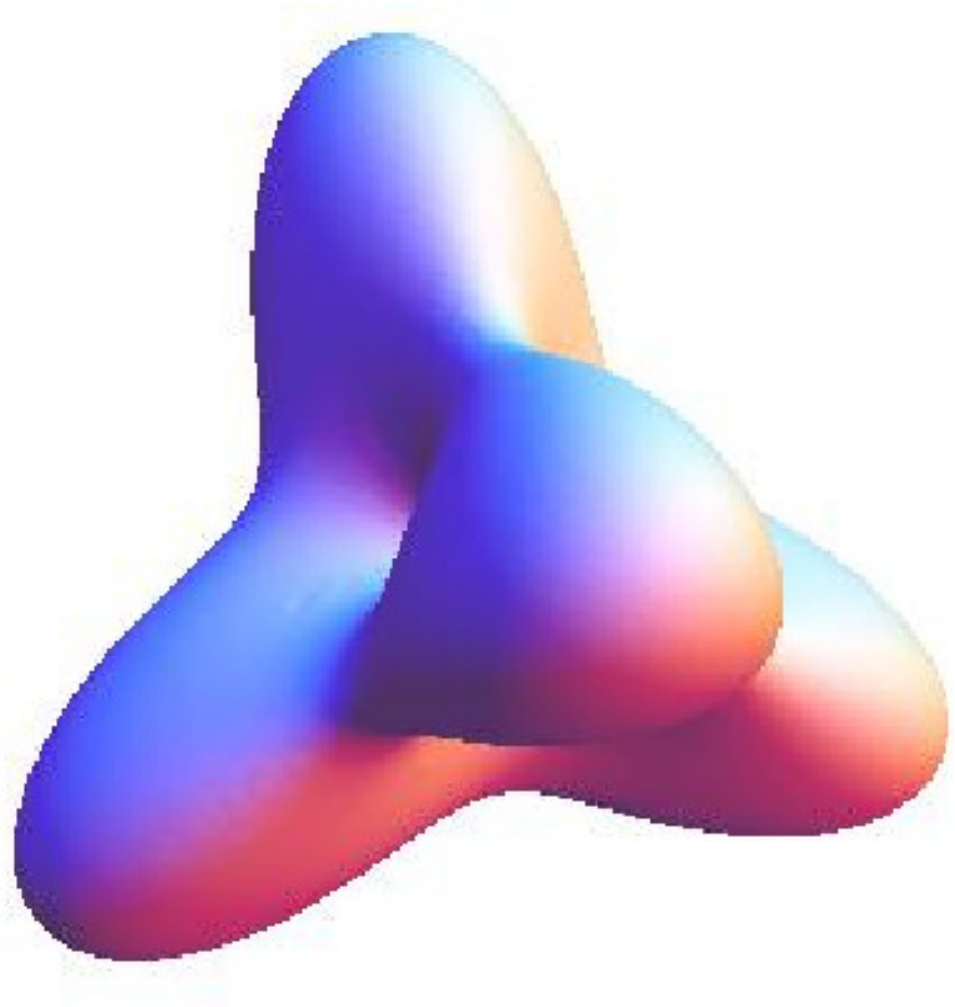}}
\end{minipage}
%\    \ \hfill
\begin{minipage}{5cm}
\subfigure[ $B=4$]{\includegraphics[scale=0.2]{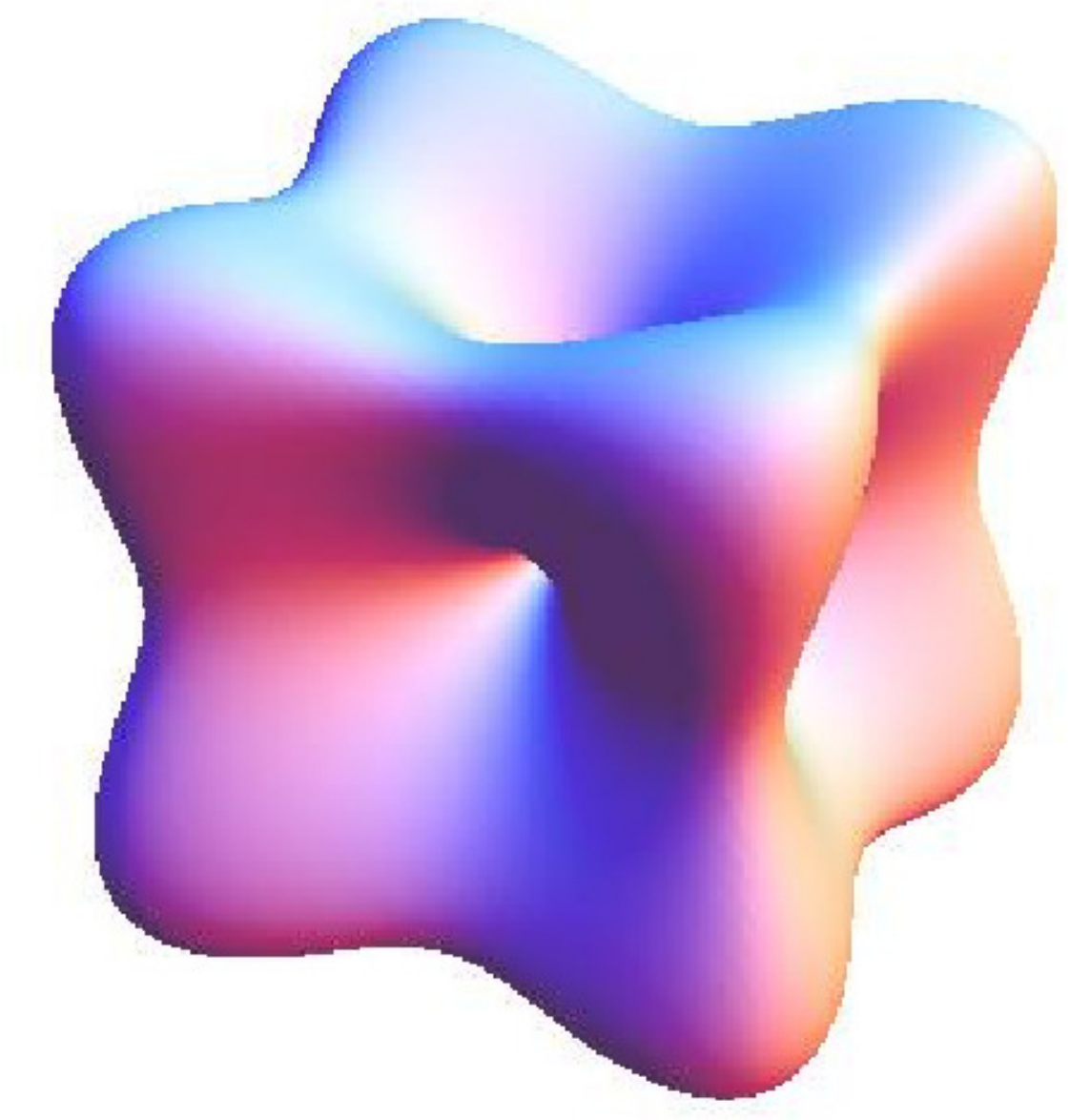}}
\end{minipage}
\    \ \vfill
%\centering{
\begin{minipage}{5cm}
\subfigure[ $B=5$]{\includegraphics[scale=0.26]{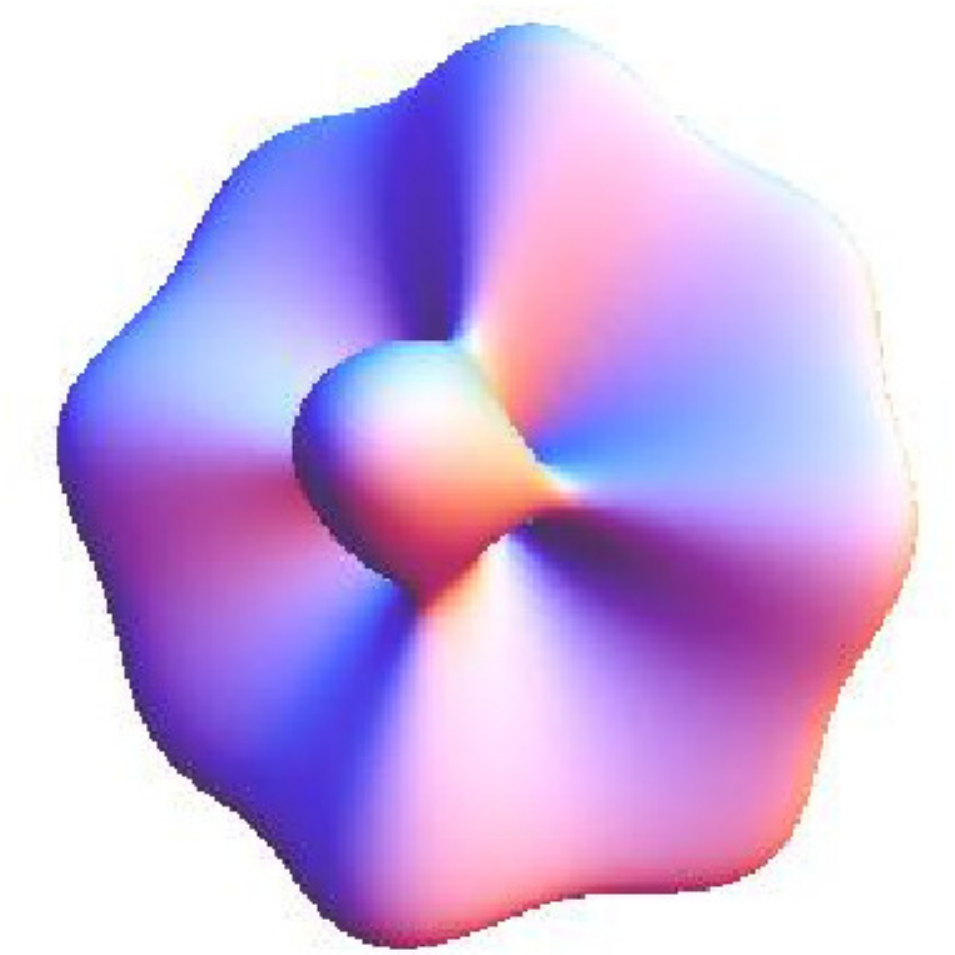}}
\end{minipage}
%\    \ \hfill
\begin{minipage}{5cm}
\subfigure[ $B=6$]{\includegraphics[scale=0.24]{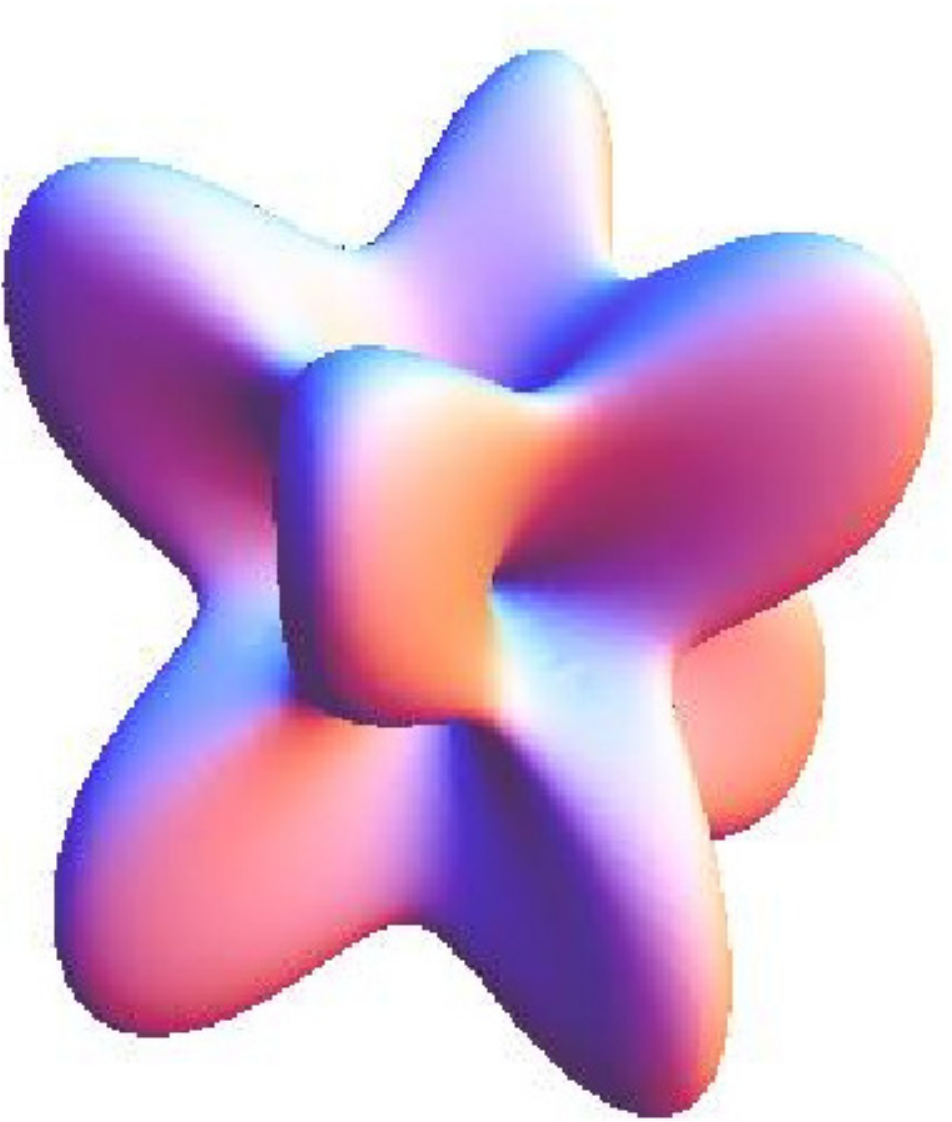}}
\end{minipage}
%\    \ \hfill
\begin{minipage}{5cm}
\subfigure[ $B=7$]{\includegraphics[scale=0.27]{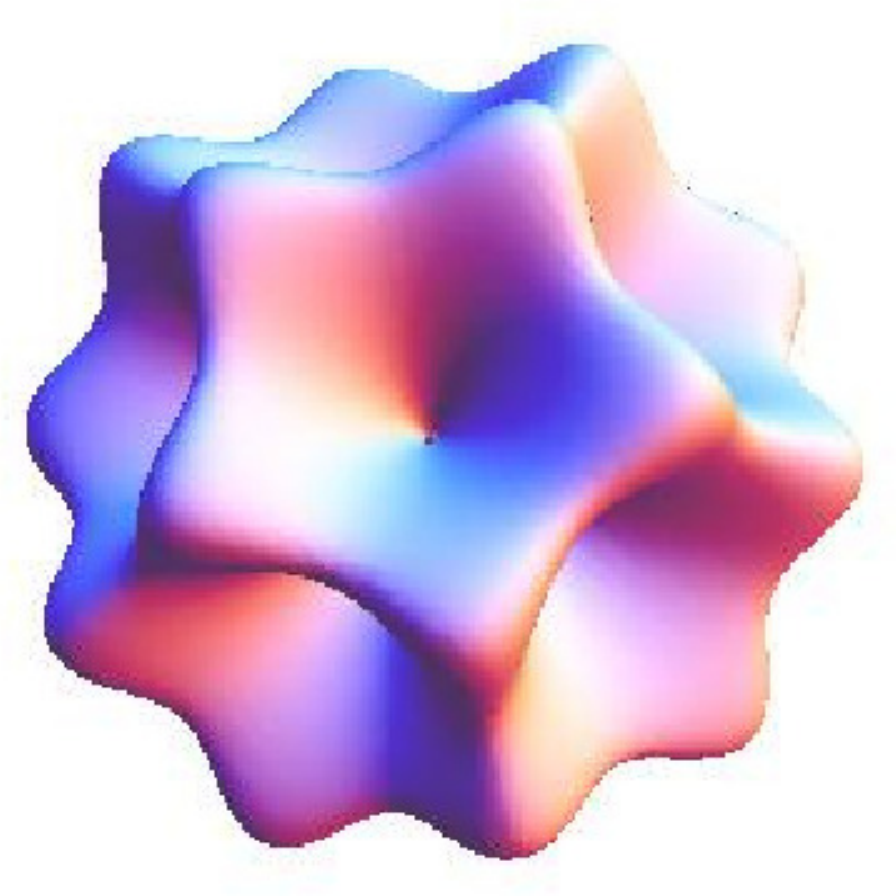}}
\end{minipage}
\    \ \vfill
%\centering{
\begin{minipage}{5cm}
\subfigure[ $B=8$]{\includegraphics[scale=0.23]{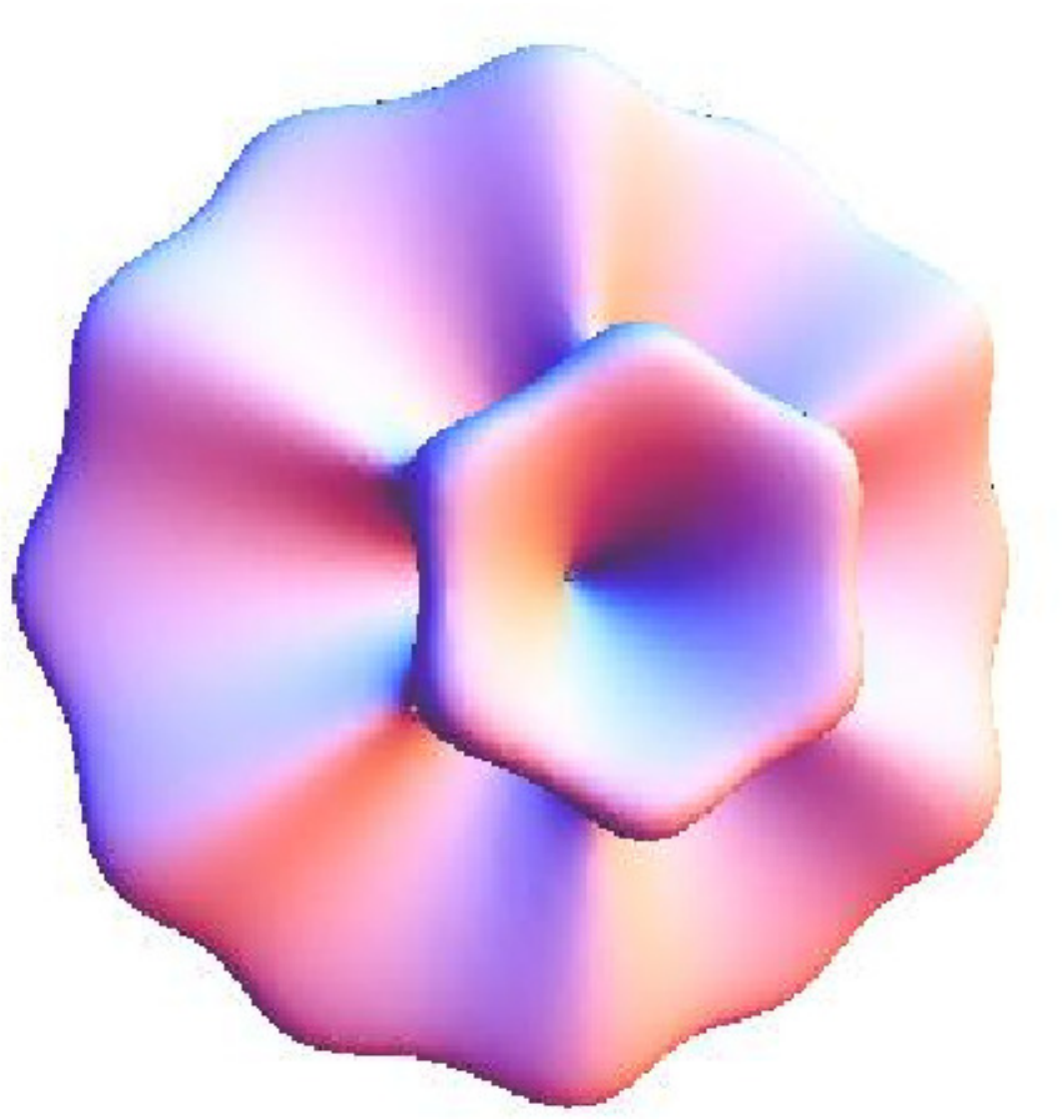}}
\end{minipage}
%\    \ \hfill
\begin{minipage}{5cm}
\subfigure[ $B=9$]{\includegraphics[scale=0.26]{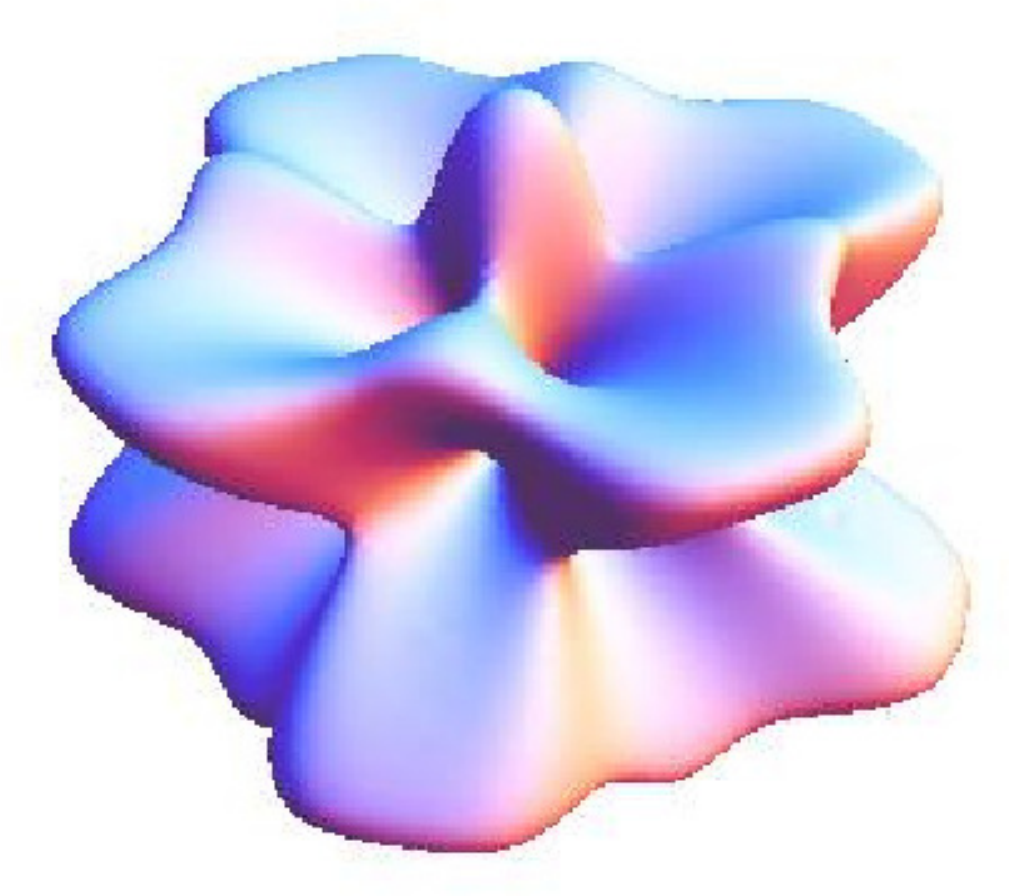}}
\end{minipage}
%\    \ \hfill
\begin{minipage}{5cm}
\subfigure[ $B=10$]{\includegraphics[scale=0.24]{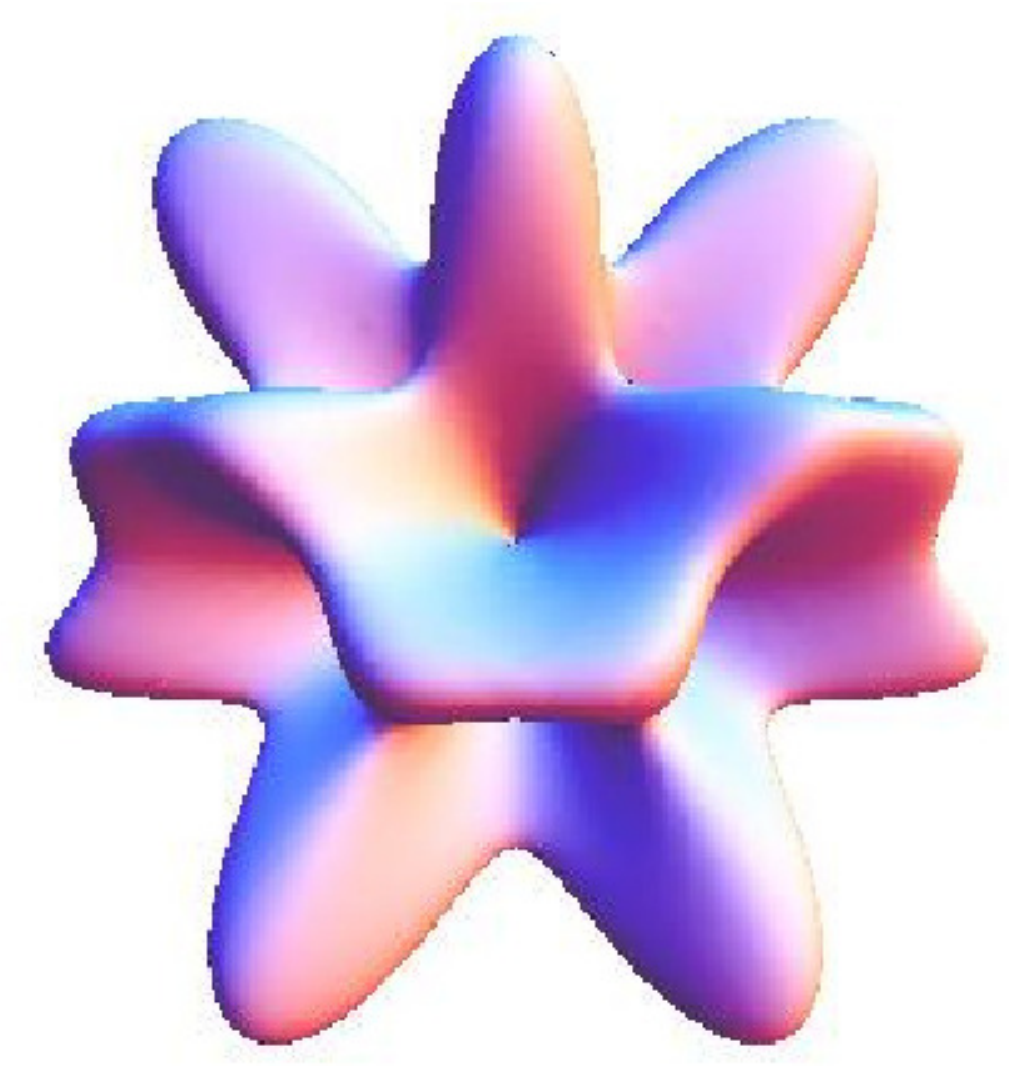}}
\end{minipage}
\    \ \vfill
%\centering{
\begin{minipage}{5cm}
\subfigure[ $B=11$]{\includegraphics[scale=0.27]{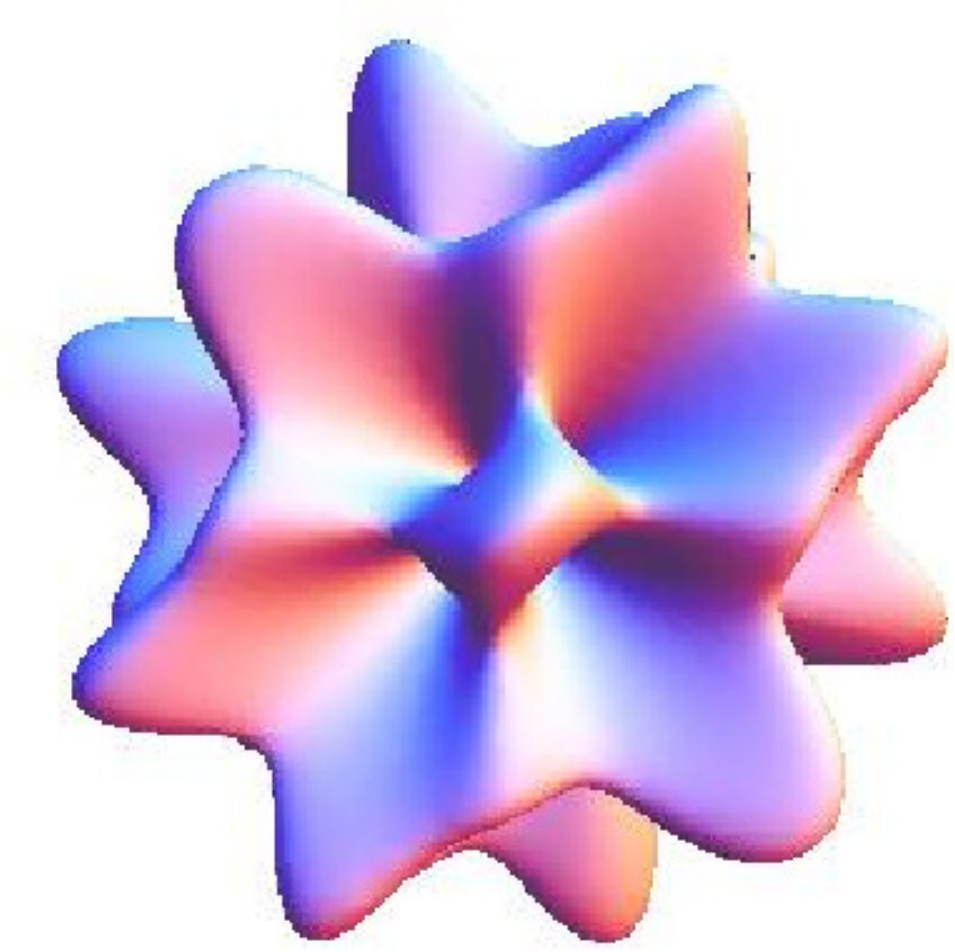}}
\end{minipage}
%\    \ \hfill
\begin{minipage}{5cm}
\subfigure[ $B=12$]{\includegraphics[scale=0.25]{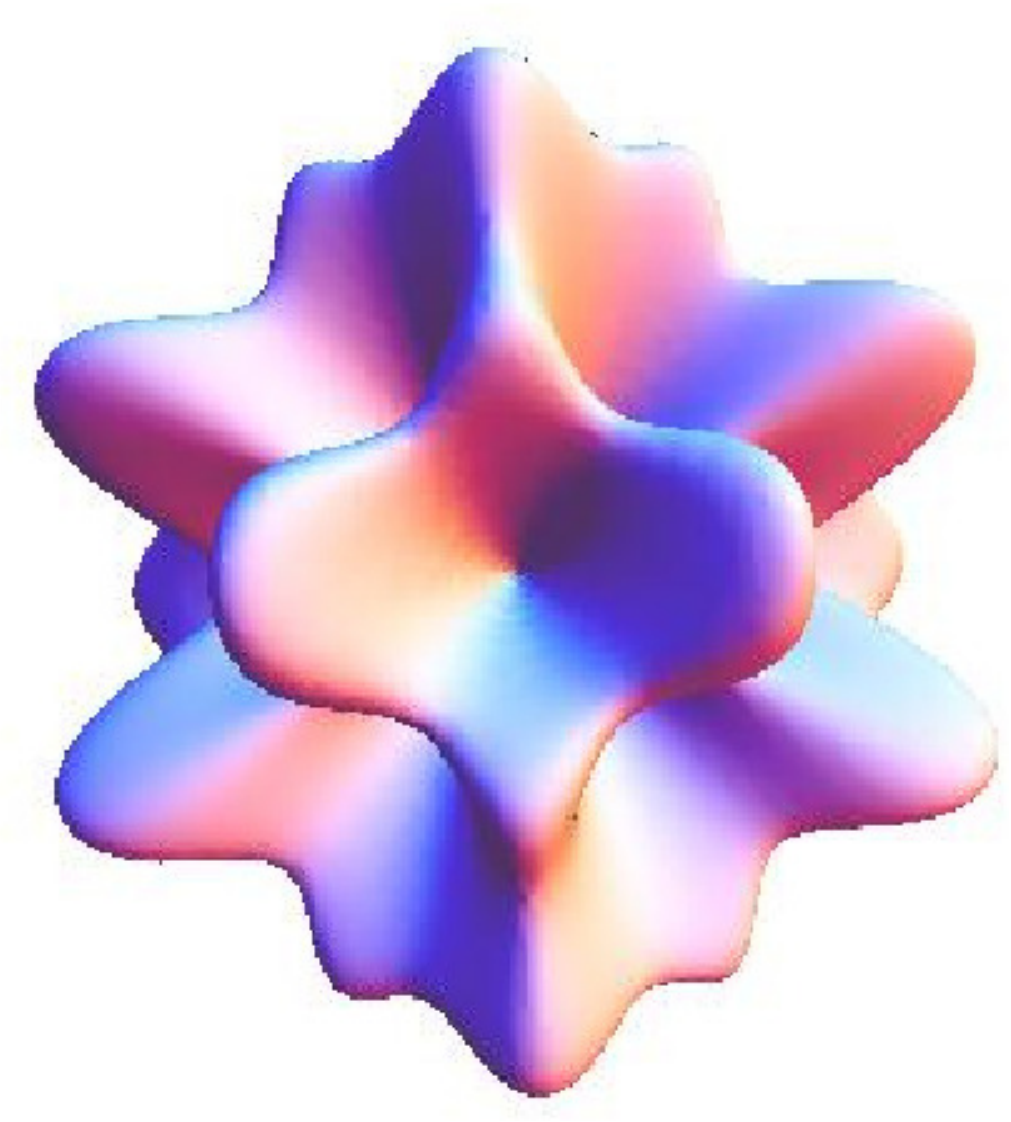}}
\end{minipage}
%\    \ \hfill
\begin{minipage}{5cm}
\subfigure[ $B=13$]{\includegraphics[scale=0.28]{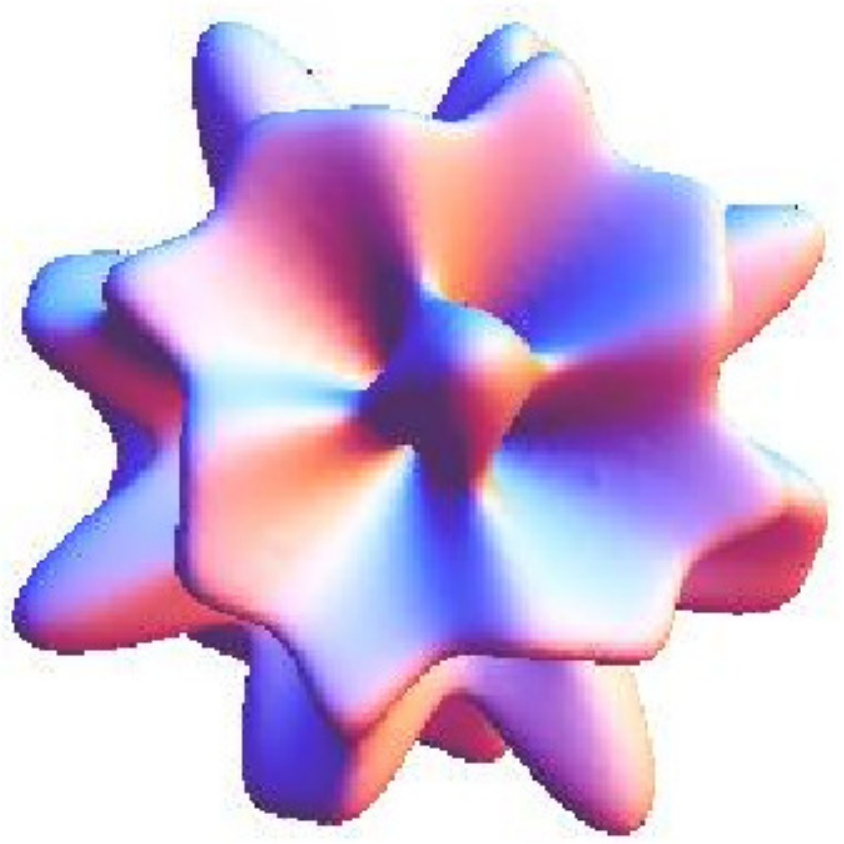}}
\end{minipage}
\    \ \vfill
%\centering{
\begin{minipage}{5cm}
\subfigure[ $B=14$]{\includegraphics[scale=0.28]{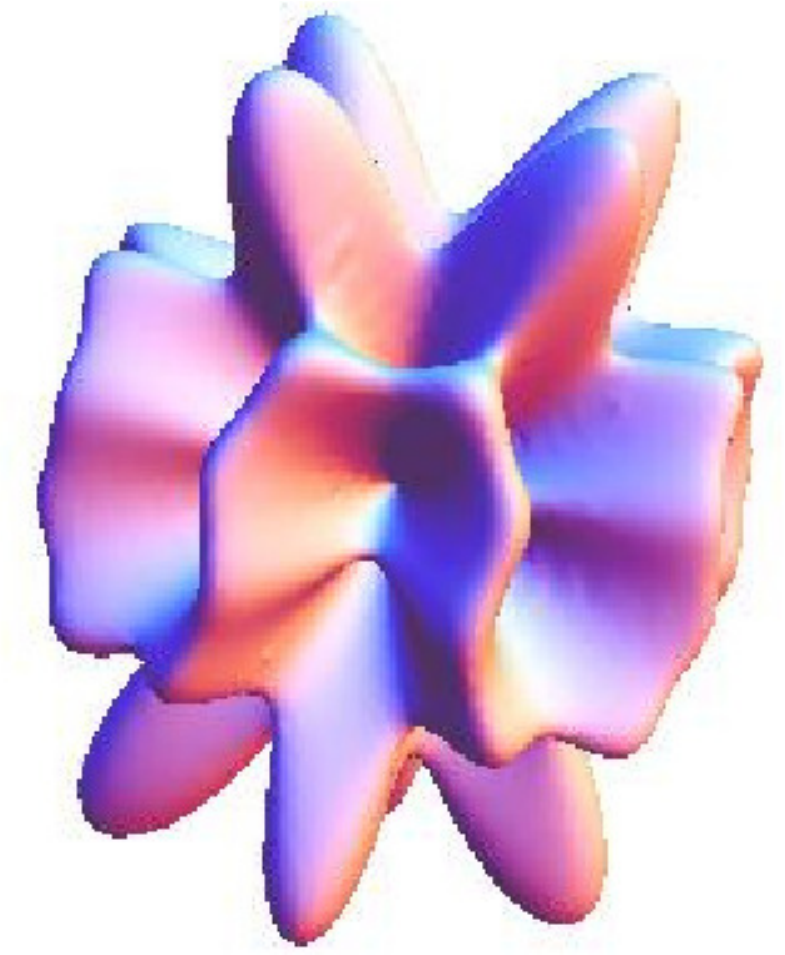}}
\end{minipage}
%\    \ \hfill
\begin{minipage}{5cm}
\subfigure[ $B=15$]{\includegraphics[scale=0.28]{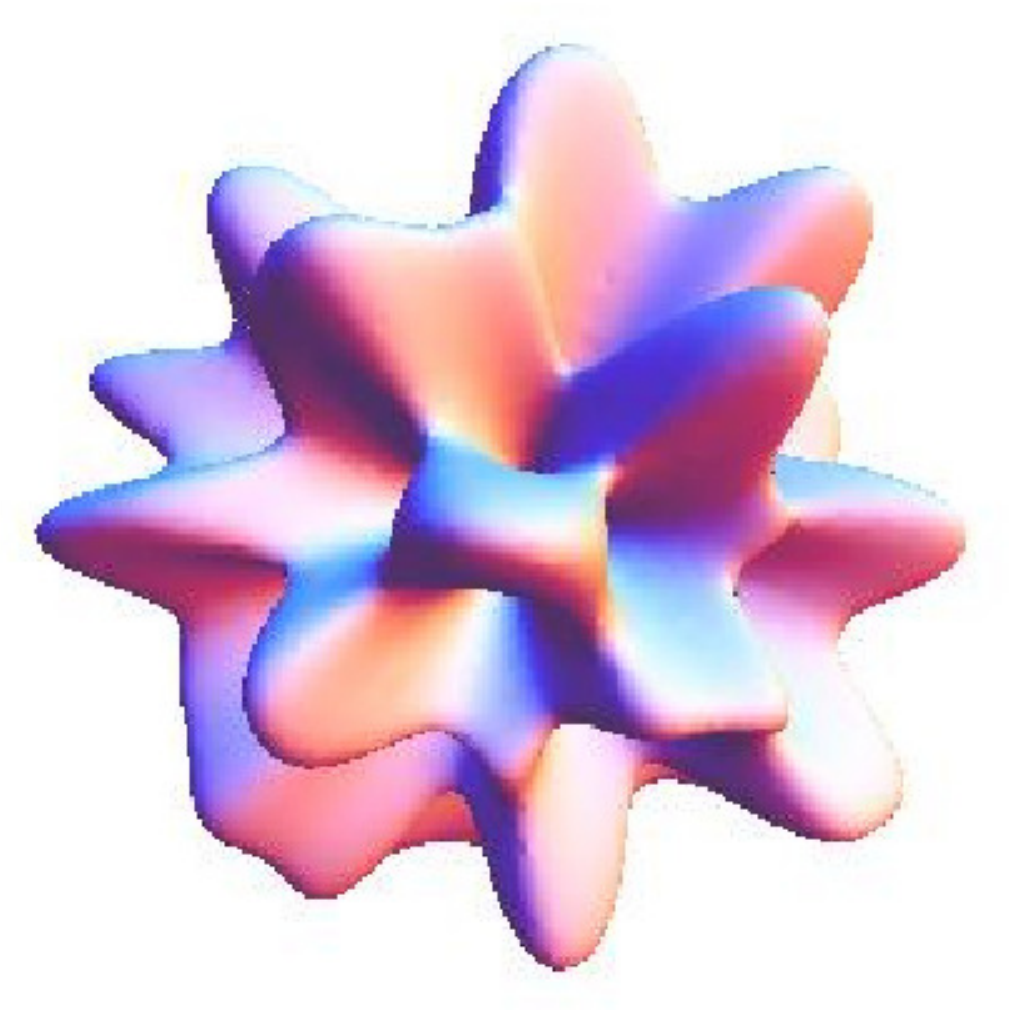}}
\end{minipage}
%\    \ \hfill
\begin{minipage}{5cm}
\subfigure[ $B=16$]{\includegraphics[scale=0.29]{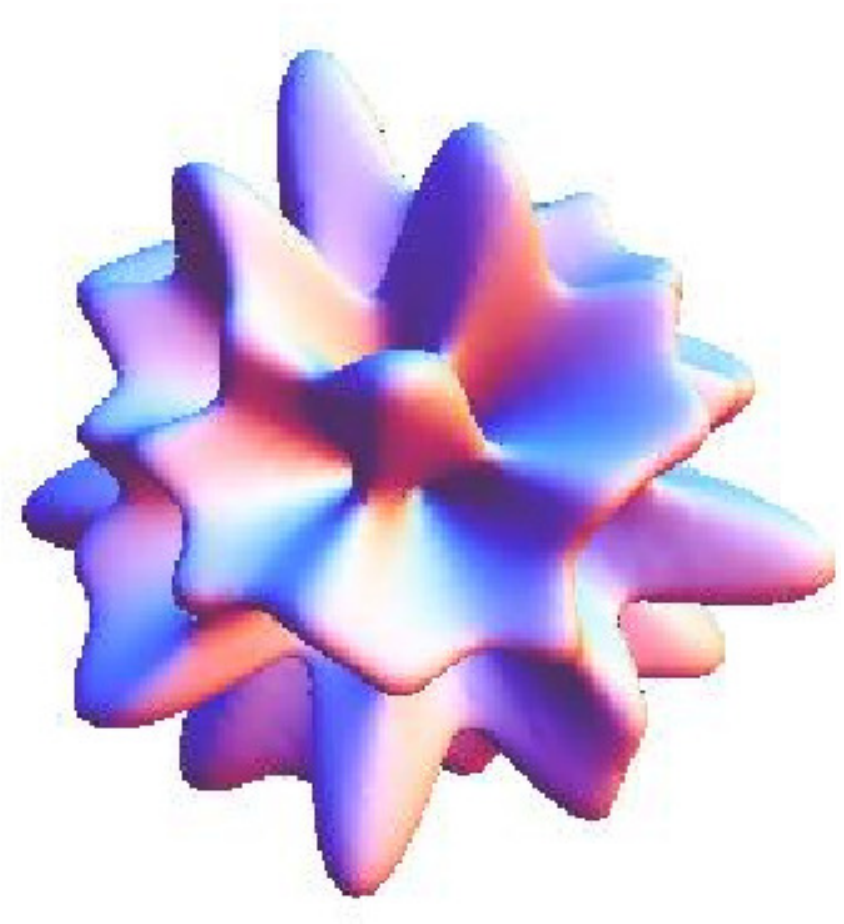}}
\end{minipage}
%\    \ \vfill
 \caption{ $V=0$. Multi-solitons for charges $ 2\leq B \leq 16 $. ($g=18$ ; $M\approx6.4 \rightarrow \kappa^2= 0.05$)}
 \label{B>2}
\    \ \vfill
\    \ \vfill
\end{figure*}

We present in figure \ref{B>2} the static configurations found within our numerical scheme, for charges $ 2\leq B \leq 16 $, for the 
vector meson stabilization of the sigma model. Our plots are incredibly similar to those on \cite{ Hen} and show once again the great
resemblance of the two models.
The calculations were done for several values of the parameters $g$ and $M$, yielding qualitatively similar configurations within each 
topological sector. Indeed, quantitative similarities also follows (e.g. see table \ref{table1}).

The expressions for the static energy of the two models are,
\begin{eqnarray}
E^{\text{\scriptsize{B-S}}}_{\text{\tiny{static}}} & = & \frac{1}{2} \int_{S^2} \epsilon \left[ (\nabla\phi)^2 + \kappa^2 (4\pi)^2 (B^{0})^2 \right] 
\label{E-BS}\\
E^{\text{\scriptsize{V-M}}}_{\text{\tiny{static}}} & = & \frac{1}{2} \int_{S^2} \epsilon \left[ (\nabla\phi)^2 + \frac{g}{M^{2}} B^{0} \Delta \omega + \frac{g^2}{M^2} (B^{0})^2 \right] 
\label{E-VM}
\end{eqnarray}
In \cite{ Sutcliffe-2d} was suggested an approximation of the solutions of \eqref{est-w} by applying a derivative 
expansion, in which at first order the Laplacian is neglected. So the energies in the two models could be compared upon an identification
of the parameters as $\kappa = \frac{g}{4\pi M}$.
Therefore, we chose our parameters $g$ and $M$ appropriately for comparison of the energies found in the present study for the vector meson 
theory with the ones obtained in \cite{ Hen} for the baby-skyrme model with $\kappa^2 = 0.05 $.
The energies (normalized with $4\pi B$) for both models are listed in Table \ref{table1}, for charges up to $B=14$.

\squeezetable
\begin{table}%[h]
\caption{ Energies and Symmetries of the vector meson and baby-skyrme models for $\kappa^2 = (\frac{g}{4\pi M})^2= 0.05$ (with $g=18$, $M\approx6.4$). 
*data taken from \cite{ Hen}}
\centering {
\begin{tabular}{c | c | c | c | c}         \hline \hline
Charge & \multicolumn{2}{c|}{Baby Skyrme *} & \multicolumn{2}{c}{Vector Meson}   \\ \hline
B      &  $E/4\pi B$ & Symmetry  & $E/4\pi B$  & Symmetry \\ \hline
2      & 1.071       & $D_{\infty h}$ & 1.068  & $D_{\infty h}$\\
3      & 1.105       & $T_d$          & 1.099  & $T_d$    \\
4      & 1.125       & $O_h$          & 1.117  & $O_h$  \\ 
5      & 1.168       & $D_{2d}$       & 1.155  & $D_{4d}$ \\
6      & 1.194       & $D_{4d}$       & 1.179  & $D_{4d}$  \\
7      & 1.209       & $I_h$          & 1.193  & $I_h$ \\
8      & 1.250       & $D_{6d}$       & 1.229  & $D_{6d}$  \\
9      & 1.281       & $D_{4d}$       & 1.256  & $D_{4d}$    \\
10     & 1.306       & $D_{4d}$       & 1.278  & $D_{4d}$ \\
11     & 1.337       & $D_{3h}$       & 1.306  & $D_{3h}$  \\  
12     & 1.360       & $T_d$          & 1.328  & $T_d$  \\
13     & 1.386       & $O_h$          & 1.352  & $O_h$ \\
14     & 1.421       & $D_2$          & 1.383  & $D_{4d}$\\  \hline \hline
 
\end{tabular}
}
\label{table1}
\end{table}

It can be seen from Table \ref{table1} that the energies are very similar in both models, being the energy on the vector meson theory only slightly 
smaller than the ones found in \cite{ Hen} for the baby-Skyrme case.\\
The symmetries we find in our solutions generally agrees with those in \cite{ Hen}, with only a minor discrepancy in the $B=5$ sector
and a more significant one on the $B=14$ solution, where we have encountered a more symmetrical configuration. It is difficult to know whether
these $B=14$ solutions are different because different models are being compared, or because one of them is only a local minima of the energy.

We note here that the symmetries of the multi-soliton solutions we have found, match exactly (up to charge 13) with the symmetry group $H_B$ of
$2B - 2$ point Coulomb charges on a sphere minimizing the energy, known as the Thomson problem.\\ (For more details see Ref. \cite{ Icosahedral}, 
in which these groups were listed). 

A final comment on Table \ref{table1} is necessary. We have fixed the ratio $\frac{g}{M}$ for comparison between the two models, as mentioned. 
Hence, there is still a freedom remaining in setting the values  $g$ and $M$ independently, and we have observed that the energies of the two models approach each other for larger 
values of both parameters, being significantly smaller in the vector meson theory when small values are considered.\\
The fact they approach for large values of $g$ and $M$ seems reasonable in view of equation \eqref{E-VM}, and the smaller values 
attained when $g$ and $M$ are small, suggest the possibility of the second term of this energy expression taking negative values.

\subsection{$V\neq0$}

The spatial scale adopted by the solitons in the previous section with $V=0$ is the natural length scale provided by the radius of 
curvature of the sphere. The inclusion of a potential term is known to endow the solitons with a 
preferred size that depends on the parameters of the theory (and not on the geometry). So, there will be two length scales, namely the 
size of the solitons and the radius of curvature of the unit sphere. And it is expected an interesting interplay between these two scales.

However, the choice for the potential is largely arbitrary (given that any potential which contains no field derivatives would do equally 
well), and it has been seen that the form of this term has a major impact on the existence and structure of baby 
multi-skyrmions \cite{ Weidig, potentials}. 

In this first exploratory work, we shall use the potential term
\begin{equation}
  V(\vec{\phi}) = m^2 (1 - \phi_3)
\label{old} 
\end{equation}
This particular form is motivated by analogy with the one traditionally used in the three-dimensional Skyrme model, and
the suggestive name given to the parameter comes from the idea of a mass term for the pion field.
Note that the inclusion of this term breaks the O(3) symmetry of the model, leaving only the O(2) subgroup.

%\subsection{Relevant Parameters}
We have three parameters in our model, namely: the pion mass $m$; the vector meson mass $M$; and the coupling constant $g$. 
We shall identify two useful combinations $\alpha$ and $\beta$, that might help to organize the exploration through the parameter space. 
The first one, $ \alpha := \left(\frac{mg}{M}\right) $, is a dimensionless quantity representing the strength of interaction with the vector meson field.
While, $ \beta := \left(\frac{mM}{g}\right)^{-1/2} $, gives the ratio between the length scale of the solitons and 
the length scale of the target space (its curvature radius), and one might regard it as a sort of \textit{soliton density}.

%\subsection{Numerical Results}
We present below the static solutions found for the topological sectors $1 \leq B \leq 9$, obtained after the diffusion process
discussed previously, namely the evolution of the parabolic system \eqref{difusion}-\eqref{difusion-w}.\\
As the main mean of representation we have chosen baryon (or topological) density ($B^0$) plots. 
Noticing that, for all the cases we have considered, the plots associated with the energy density are very similar 
to these ones.

%\subsubsection{The first two sectors}

\begin{figure}%[ht]
\centering{
\begin{minipage}{4.2cm}
 \subfigure[]{\includegraphics[scale=0.13]{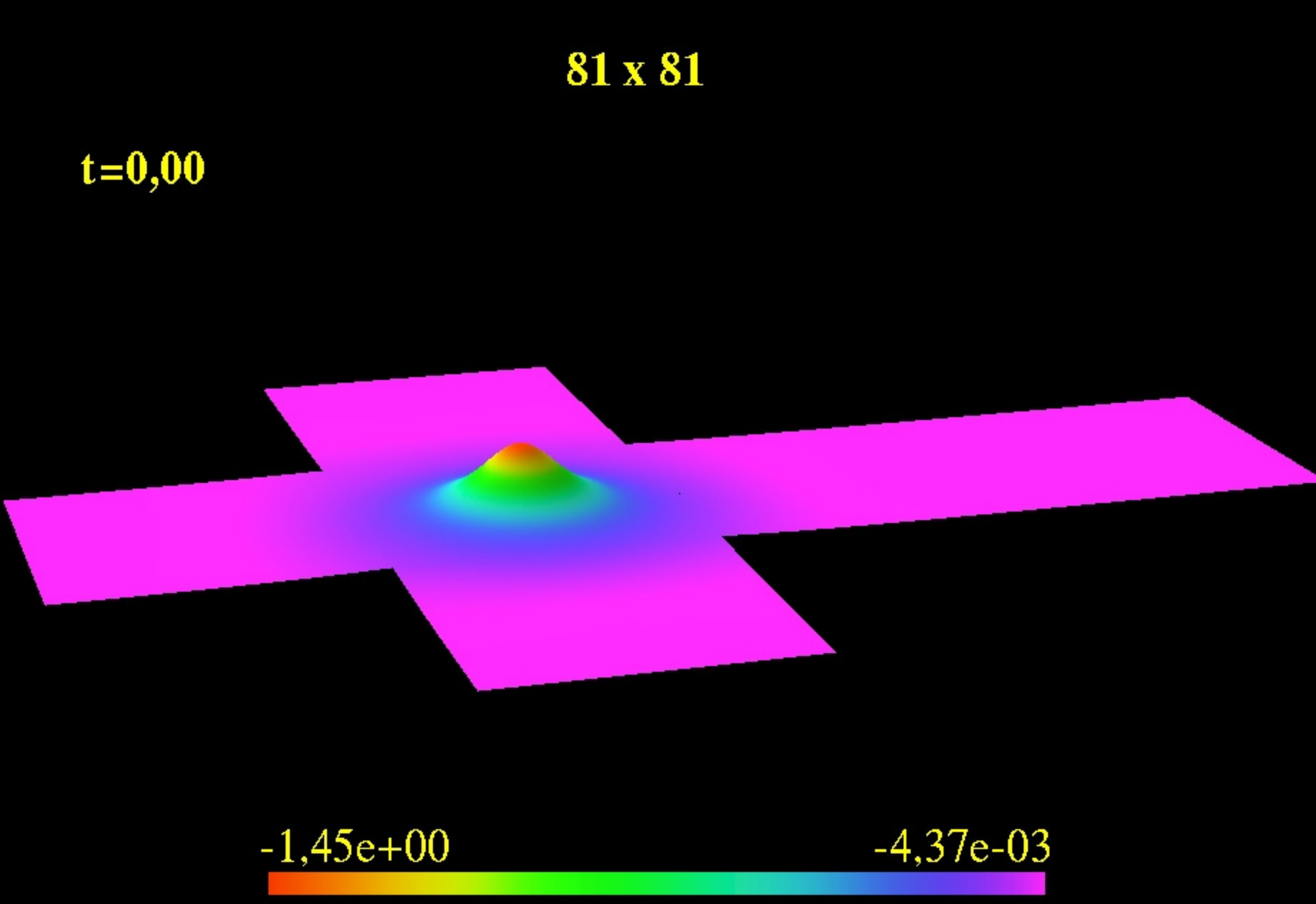}}
\end{minipage}
\begin{minipage}{4.2cm}
\subfigure[]{\includegraphics[scale=0.13]{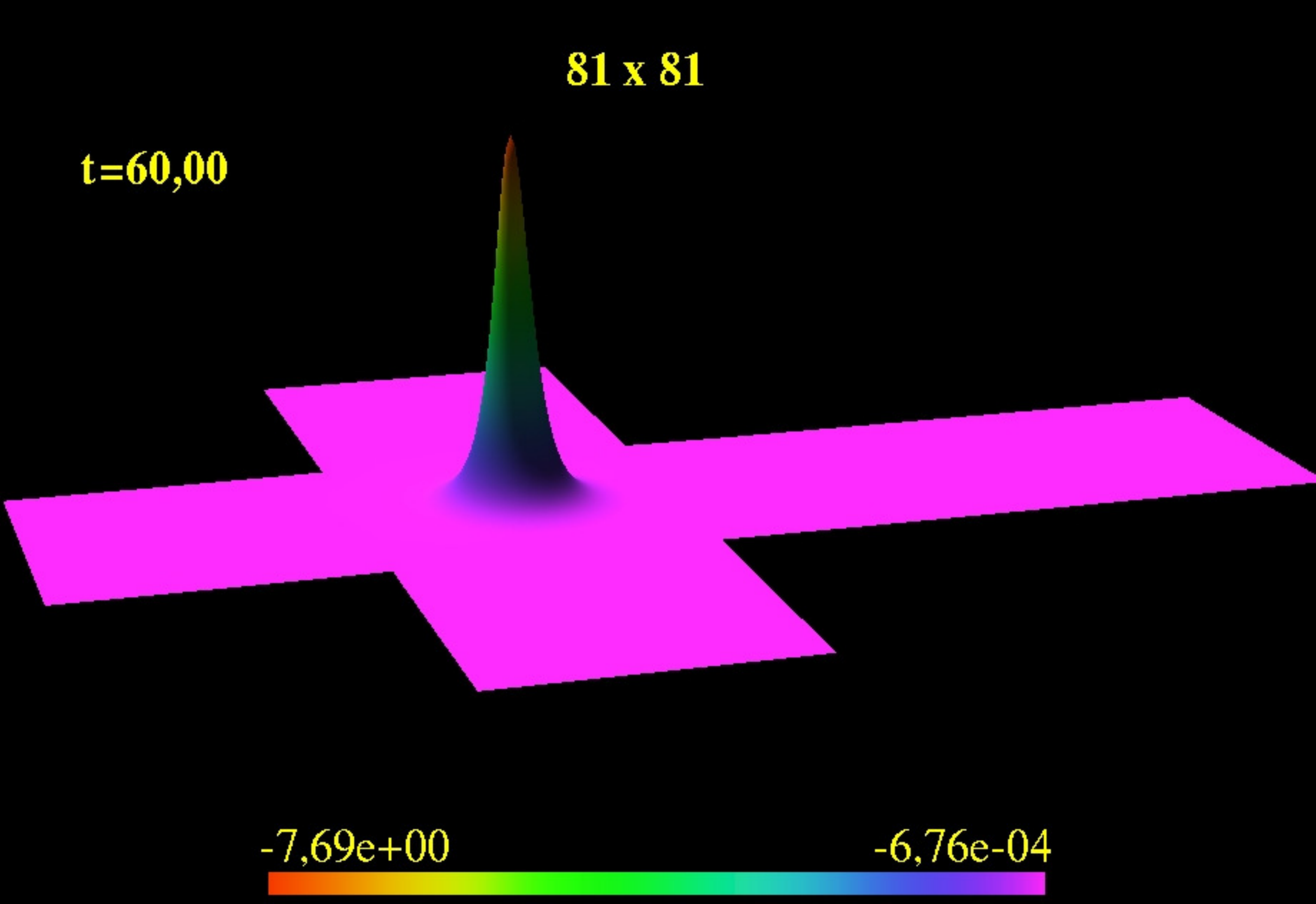}}
\end{minipage}}

\caption{ B=1. Initial and final configuration. ($g=2$, $M^{2}=16$, $m^{2}=0,1$)}
\label{fig2}
\end{figure}

In the topological sectors with charge $B=1$ and $B=2$ we have found static solutions that are axially symmetric in all the cases. They preserve the
whole O(2) subgroup.
In the sector of topological charge one, these solutions represents lumps of energy (and topological charge), whose sizes are in correspondence 
with the parameters employed. (See Fig. \ref{fig2}). 
When $\beta$ increases, the soliton spreads out in space, until it reaches a uniform distribution in the 
$\beta \rightarrow \infty$ limit, the one previously presented in the $V=0$ case.

\begin{figure}
\centering{
\begin{minipage}{4.2cm}
 \subfigure[$\text{  } \beta = 1.06$]{\includegraphics[scale=0.13]{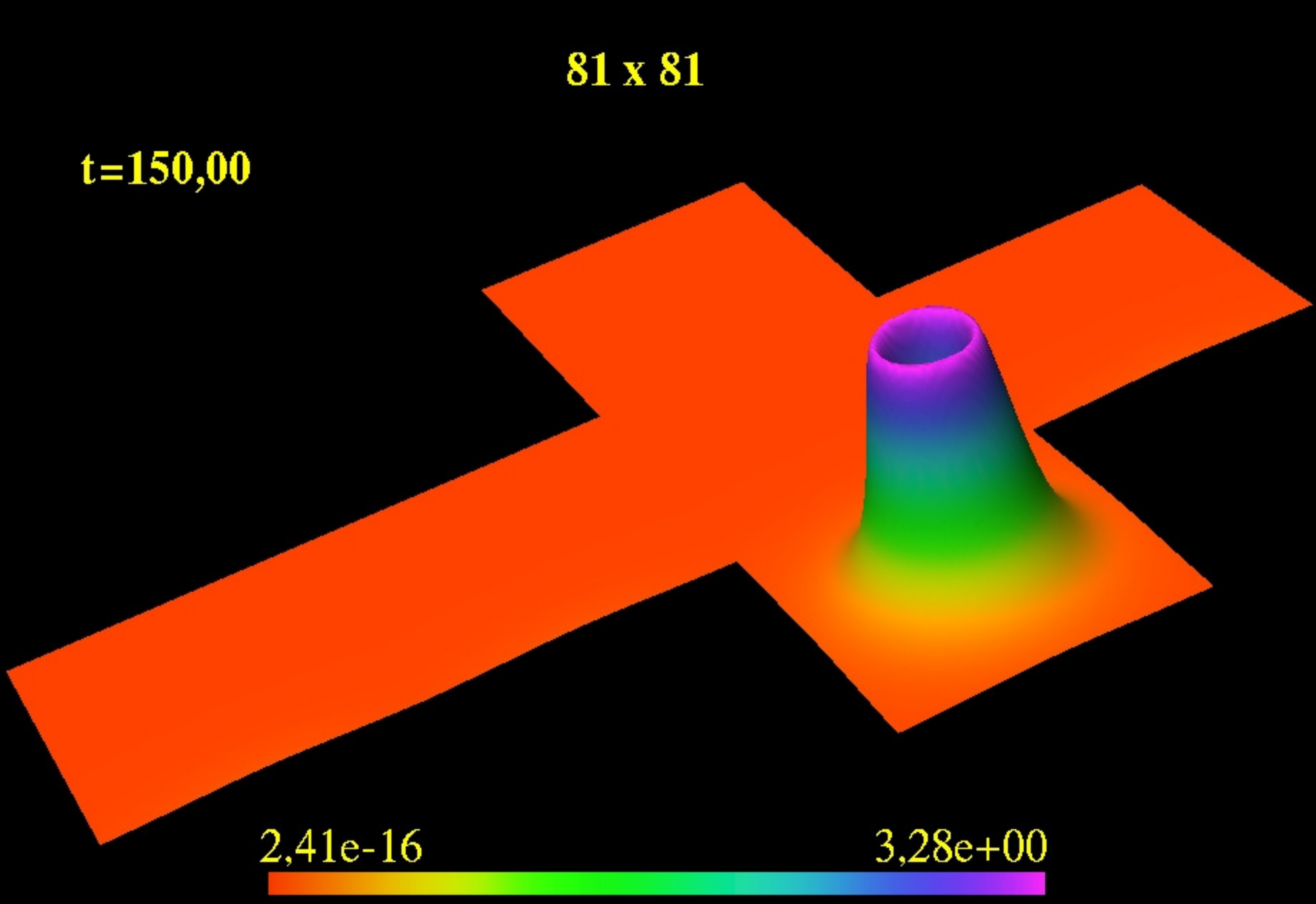}}
\end{minipage}
\begin{minipage}{4.2cm}
\subfigure[$\text{  } \beta = 3$]{\includegraphics[scale=0.13]{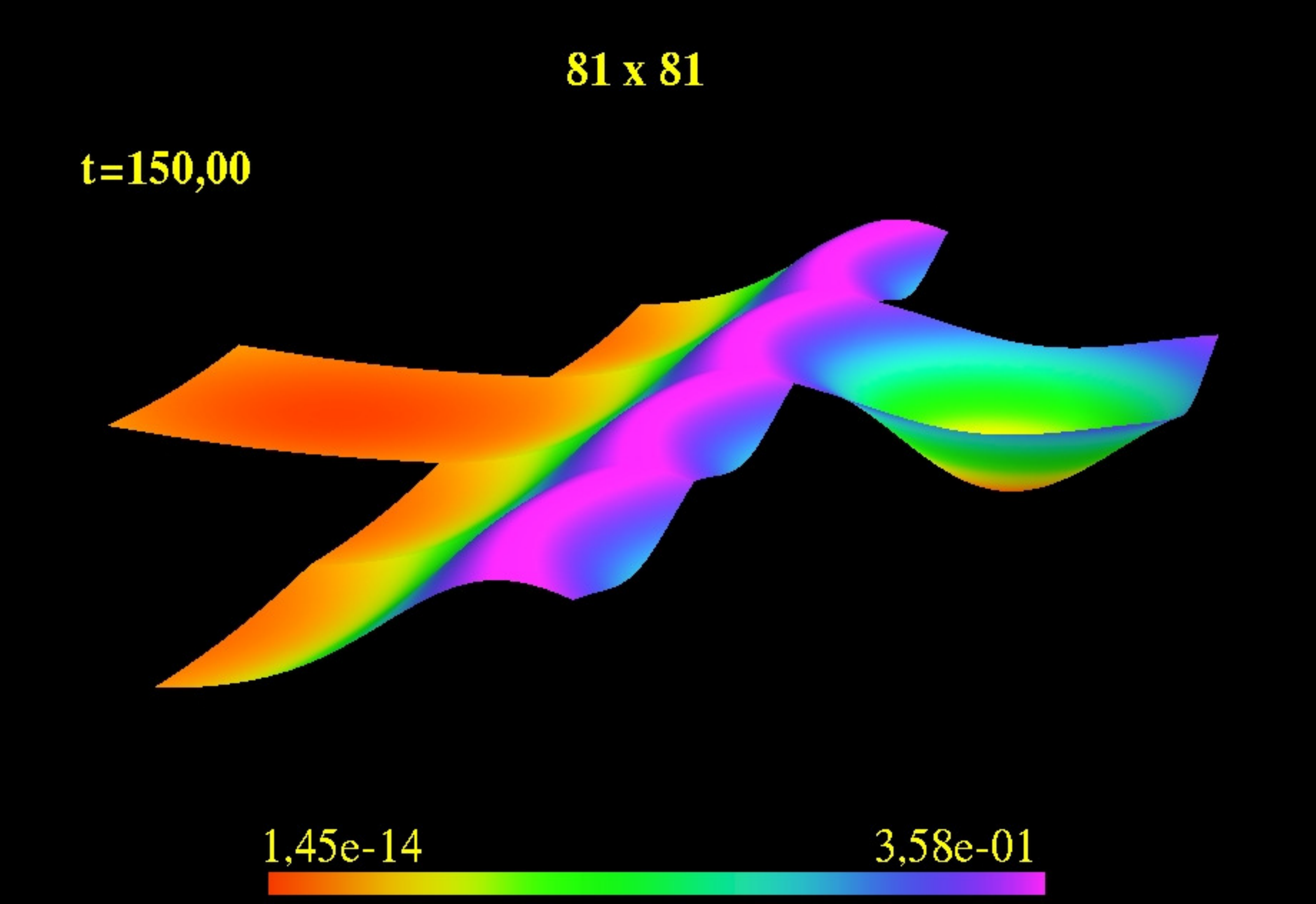}}
\end{minipage}}

\caption{ B=2. Static configurations for different densities $\beta$.}
\label{B2}
\end{figure}

In the sector $B=2$ instead, we have found ring-like configurations. The radius of these rings are of course related to the parameters, and 
again $\beta$ seems to be the most relevant among them. 
For small values of $\beta $ one sees the rings localized around one of the poles, while for large values, the radius increases and 
approaches the equator.(See Fig. \ref{B2})

All these solutions appear to be stable when introduced as initial data for the hyperbolic code in which the full dynamics of the model is 
tested, even if a perturbation is introduced into the initial set up. In the case of the rings, for example, if one introduces a small radial 
perturbation, the ring oscillates back and forth around an equilibrium position. 

%\subsubsection{Multi-Solitons}

In topological sectors of higher charges ($B>2$) we have found a rich variety of static solutions according to the  parameters used.
There are no axially symmetric configurations any more. Despite starting with axially-symmetric initial data, during the relaxation procedure 
(parabolic code) this symmetry eventually breaks down.\\ 
We have noticed that originally, this symmetry breaking took place at the interfaces of our grids structure, 
where less accuracy from the finite difference operators were expected.
Thus, in order to avoid as much as possible the influence of the grids infrastructure on the system's behavior, 
we have included a numerical perturbation into the parabolic code as explained above (see discussion in A).

The multi-solitons solutions can be arranged basically in two regimes: a low density ($\beta \lesssim 2$) and a high density ($\beta \gtrsim 3$) 
phases. The transition between these two regimes doesn't seem to be sharply defined. 
As noted in \cite{ Innocentis}, is not a phase transition in the usual sense, and this is partly due to the asymmetry of the system 
(because of the potential term used).\\
The main qualitative difference between the two regimes is that, in the first one (small $\beta$), the solitons are localized in space, grouped 
generally in pairs and individuals. % that atracts each other to form solutions with certain discrete symmetries. 
While in the high density phase (large $\beta$), the solitons are spread over the whole sphere forming very structured configurations. 

\subsubsection{Small $\beta$ regime}

Clearly, the (two-dimensional) plane case is meant to be the limit in which $\beta=0$ and $\alpha\propto g$.  
%\footnote{In the rescaling $m,M \rightarrow \infty$, but with $\frac{m}{M}\rightarrow cte$}. 
Thus in this regime of small values for $\beta$, the configurations are expected to be localized and to share many of the features present in 
the flat-space studies like \cite{ Piette-1, Weidig, potentials, chains} for 
baby-Skyrmions, or \cite{ Sutcliffe-2d} for the (2D) vector meson theory.\\ 
Although, of course, we cannot reach the limit $\beta=0$ (which would imply infinite resolution on the grids), we were able to find 
configurations localized enough to reflect these properties.

\begin{figure}
\centering
\includegraphics[scale=0.2]{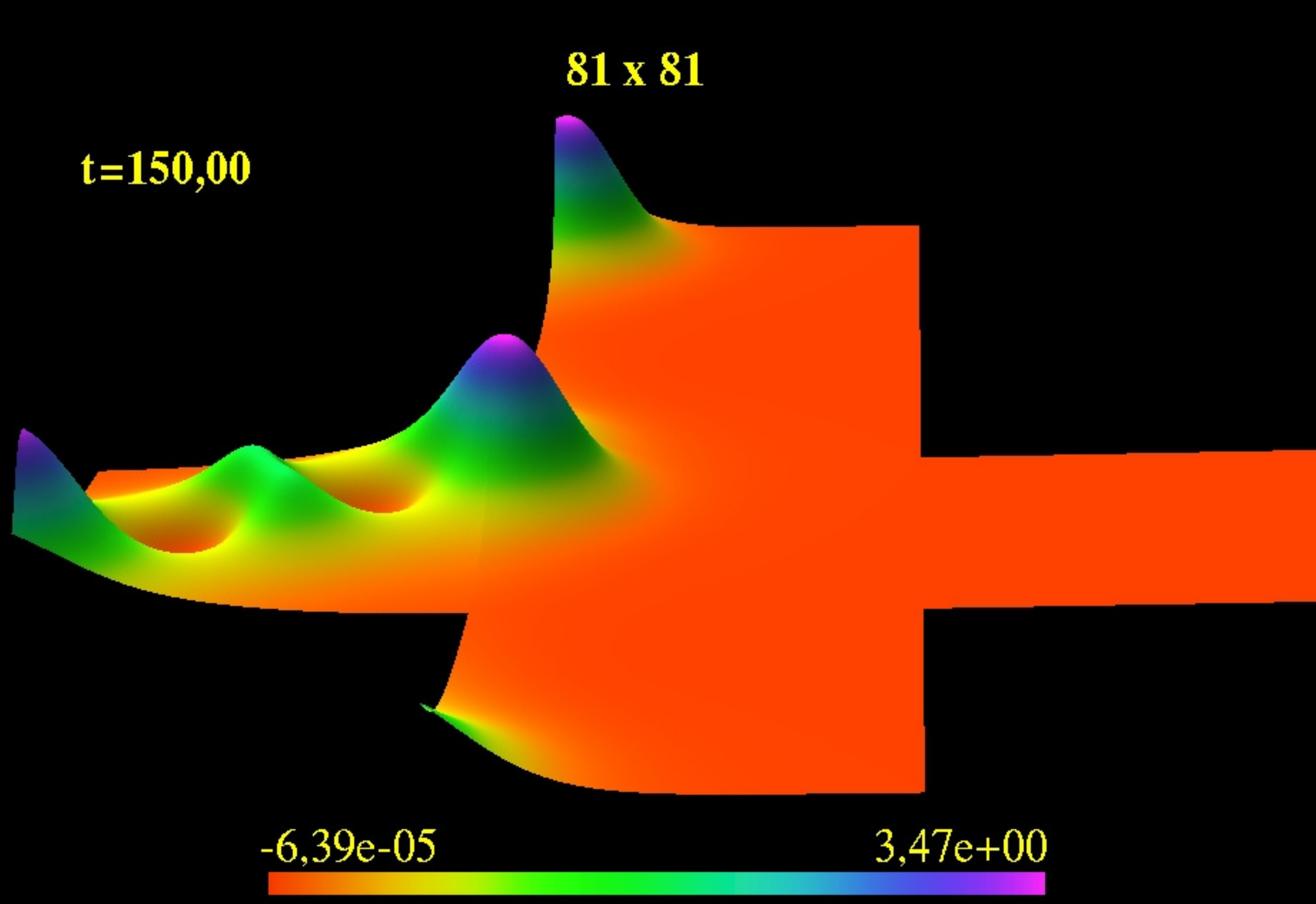}
\caption{B=3. Low density 3-soliton. ($\alpha=1.58$ ; $\beta = 1.26$).}
\label{B3}
\end{figure}

For $B=3$, our numerical procedure leads to configurations like the one displayed in figure \ref{B3}, where the solution is made up of three 
distorted solitons aligned and binded together. This is the well known 3-soliton that appears in all the studies mentioned above, including 
\cite{ Sutcliffe-2d} in which the similarities of the two models were established.

\begin{figure}

\centering{
\begin{minipage}{4.2cm}
 \subfigure[$\text{  } B=4 $]{\includegraphics[scale=0.13]{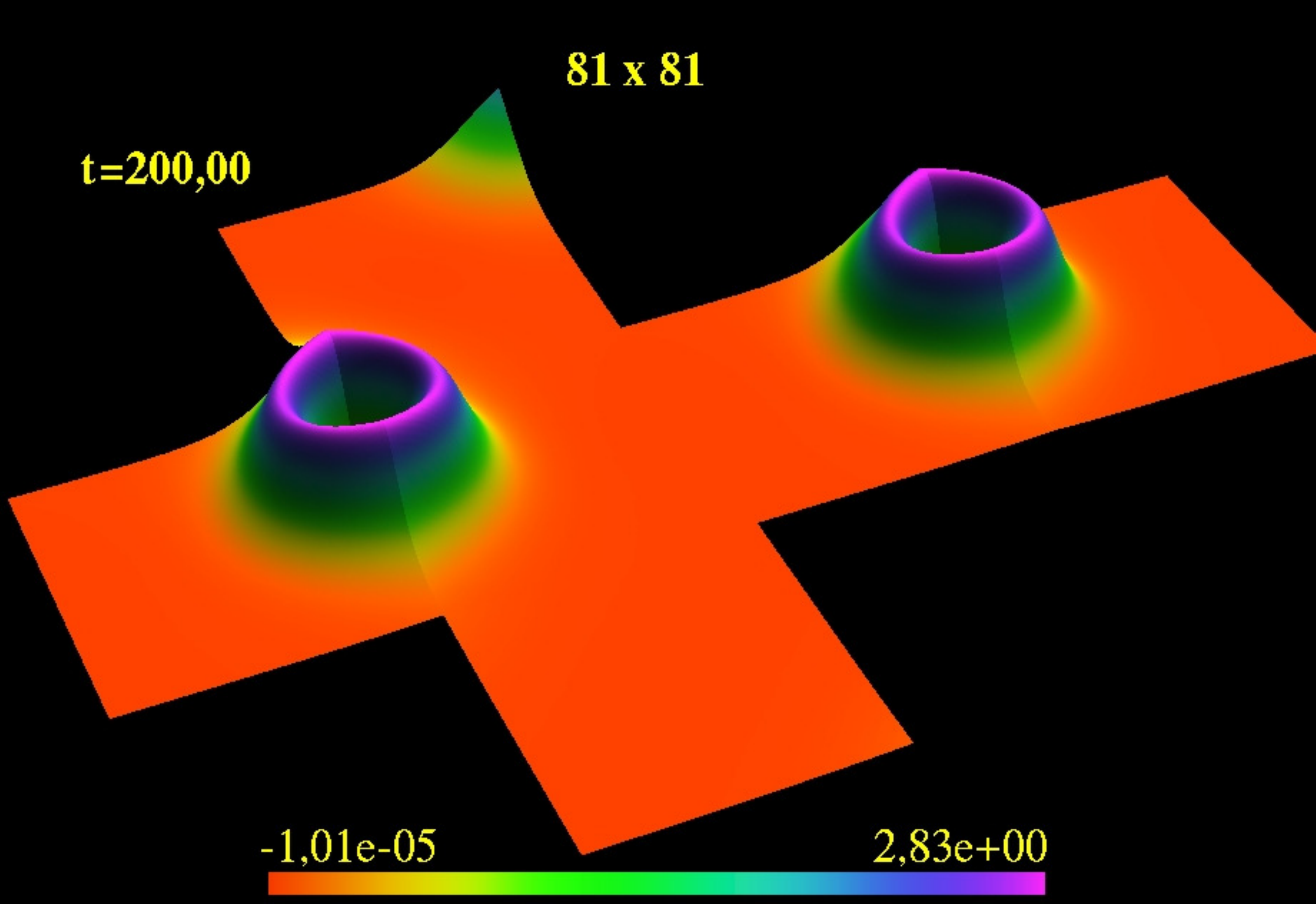}}
\end{minipage}
\begin{minipage}{4.2cm}
 \subfigure[$\text{  } B=5 $]{\includegraphics[scale=0.13]{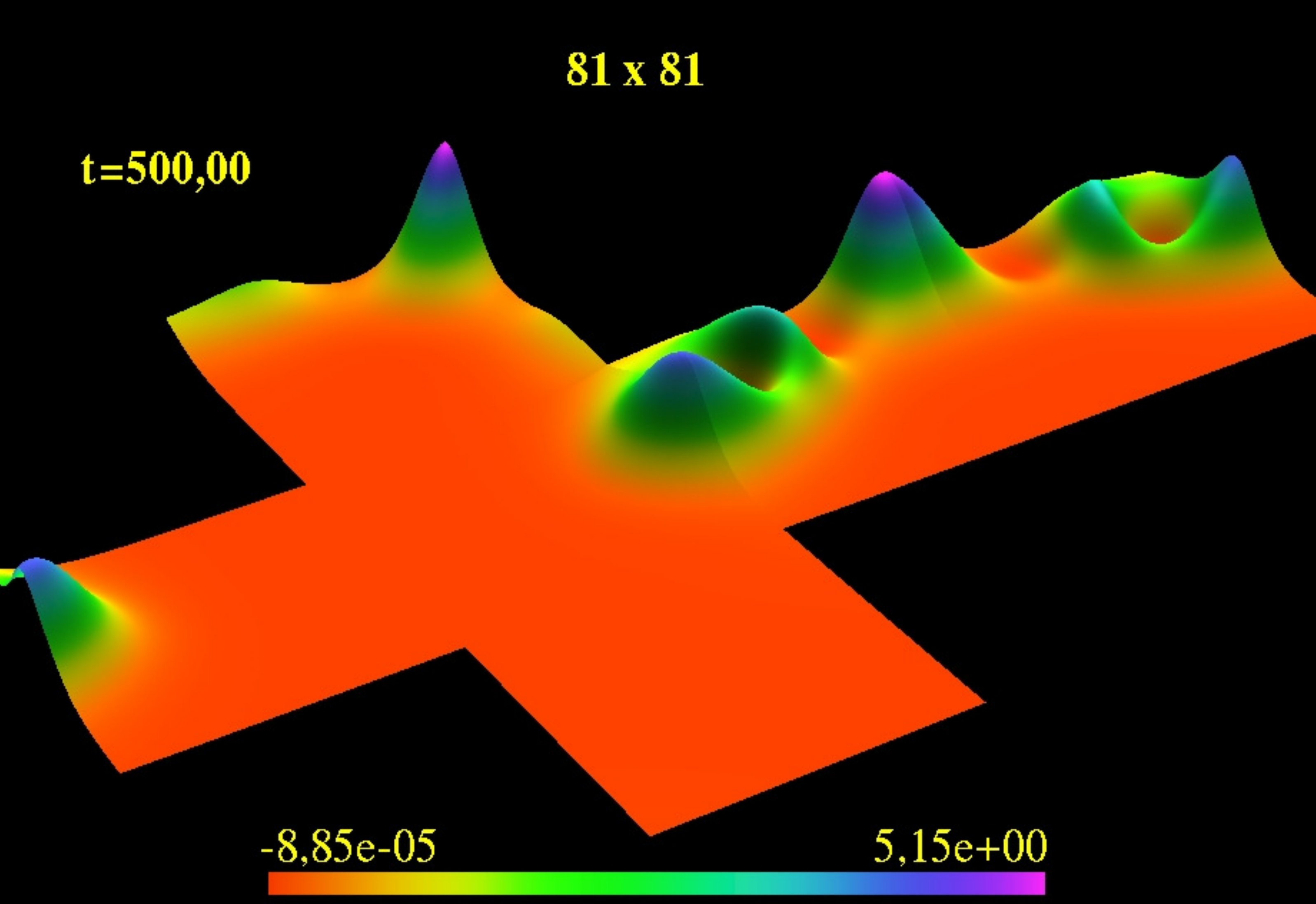}}
\end{minipage}}
\    \ \vfill
\centering{
\begin{minipage}{4.2cm}
\subfigure[$\text{  } B=6 $]{\includegraphics[scale=0.13]{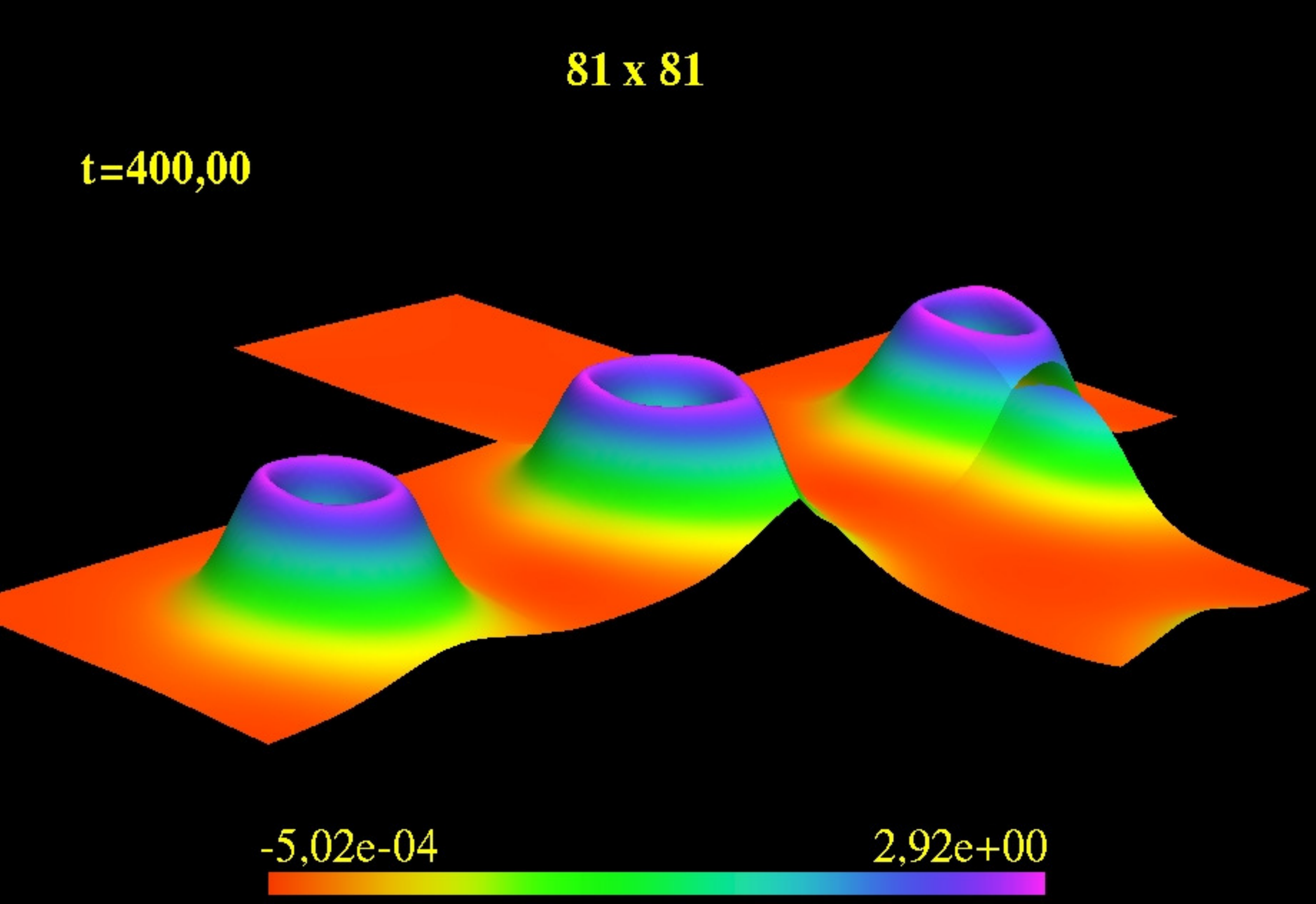}}
\end{minipage}
\begin{minipage}{4.2cm}
\subfigure[$\text{  } B=7 $]{\includegraphics[scale=0.13]{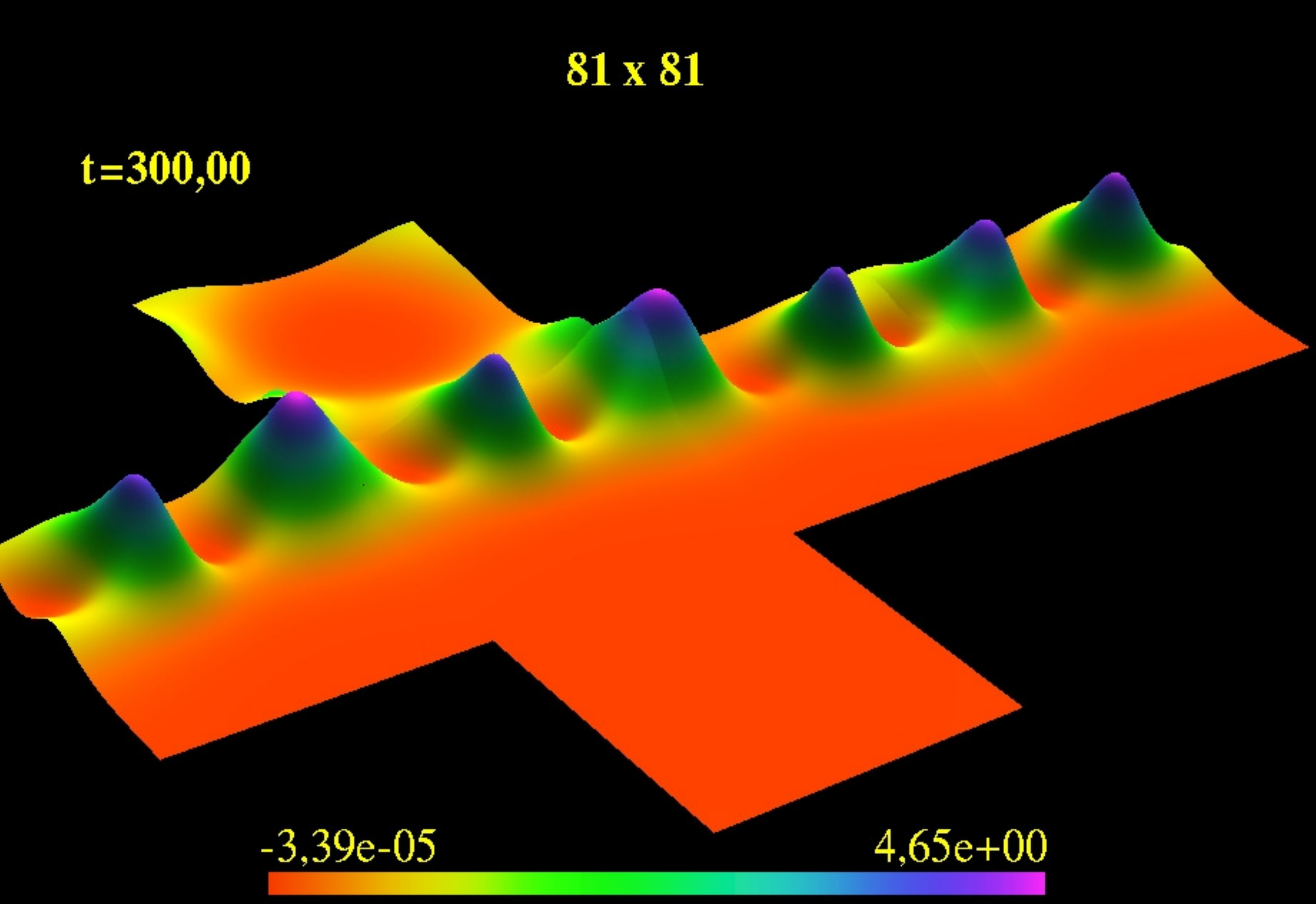}}
\end{minipage}}
\    \ \vfill
\centering{
\begin{minipage}{4.2cm}
\subfigure[$\text{  } B=8 $]{\includegraphics[scale=0.13]{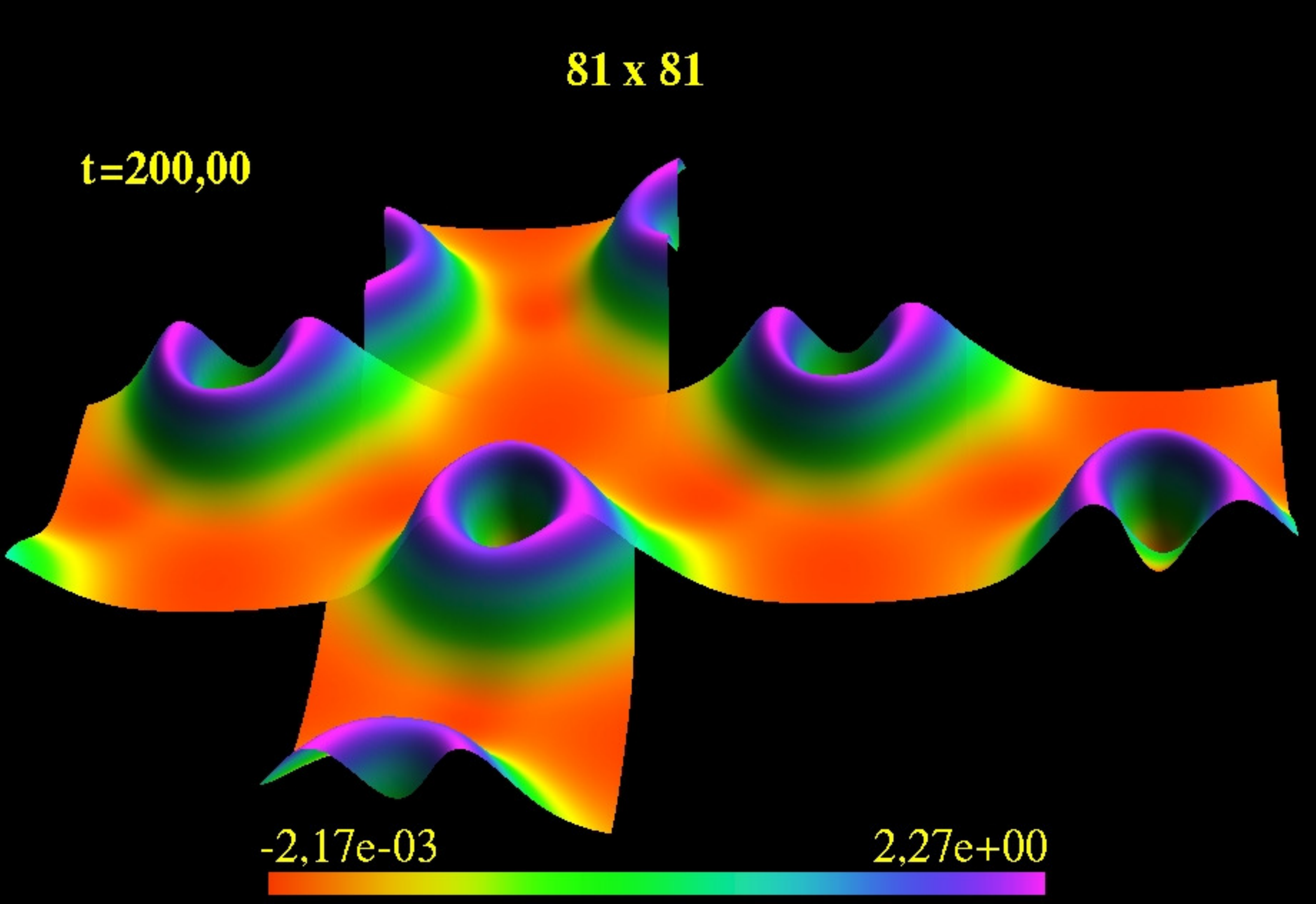}}
\end{minipage}
\begin{minipage}{4.2cm}
\subfigure[$\text{  } B=9 $]{\includegraphics[scale=0.13]{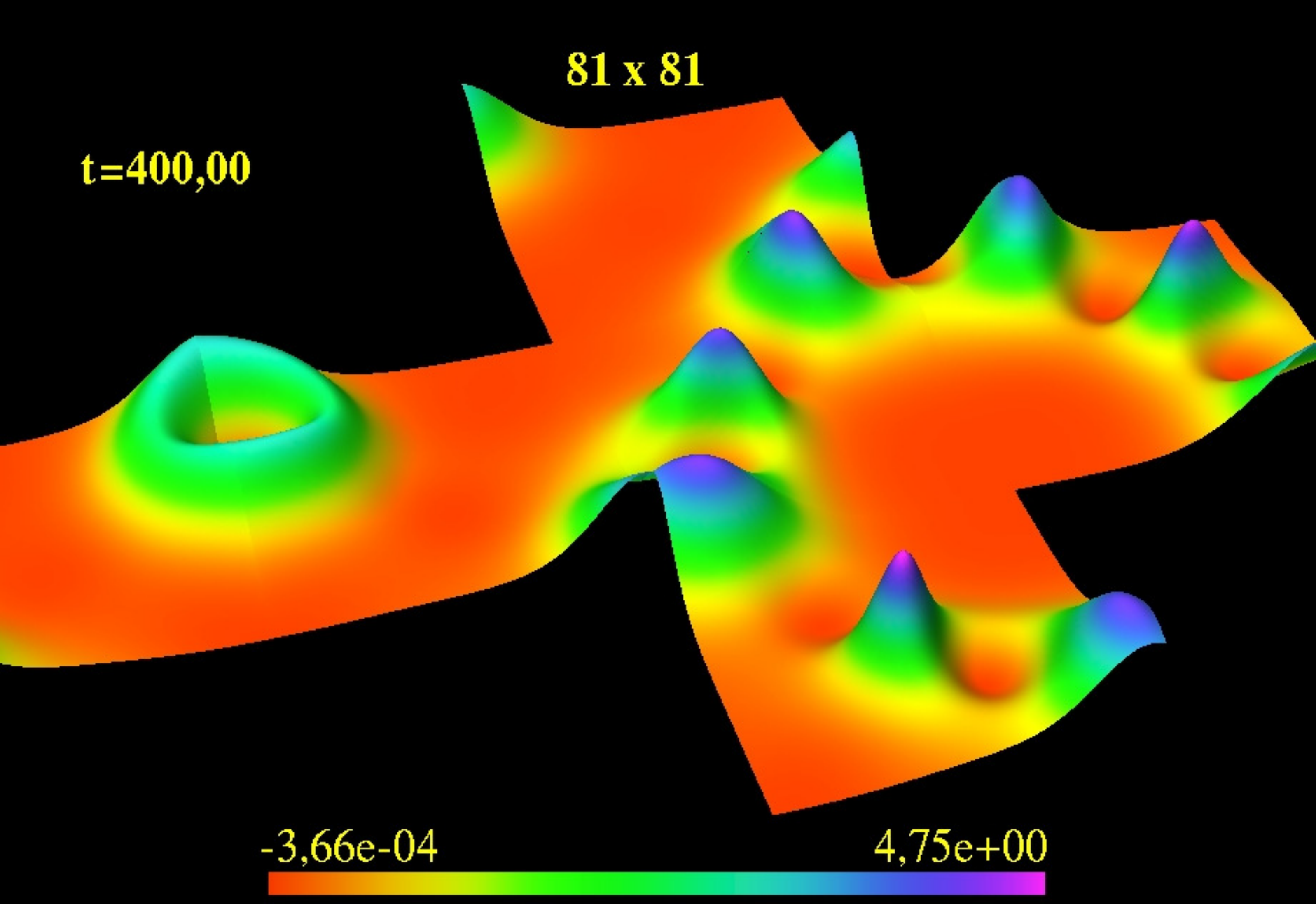}}
\end{minipage}}

\caption{ Low density multi-solitons of degee $4\leq B \leq 9$. ($\alpha=4.24$ ; $\beta = 1.03$).}
\label{B>4}
\end{figure}

For higher sectors ($B\geq4$), the baryon density plots in figure \ref{B>4} shows the configurations obtained for one
of the several parameters explored for values close to $\beta=1$. All presenting the same qualitative features.\\   
The plots suggest that the 2-soliton serves as a basic building block for higher multisolitons, especially for those sectors with an even 
topological charge.
As pointed out in \cite{ Weidig}, the 2-soliton interaction seems to be the energetically most favorable break-up mode.
We have observed this behavior during the evolution of equations \eqref{difusion}-\eqref{difusion-w} (diffusion process).
In the cases where $B$ is odd, the system breaks-up into soliton pairs and one individual soliton. 
After the break-up, the constituents start to attract each other to form structured states. However, these states appears to be very weakly 
bounded, and exist in the literature different possibles combinations of the individuals, 2-solitons and 3-solitons constituents, forming 
various crystal-like patterns within each topological sector. 
Also, there are one further type of multi-solitonic solutions, first proposed in \cite{ chains}, consisting on skyrmion chains.

Clearly, the decision of whether a given configuration is a local or a global minima of the energy becomes subtle in this context.
For that reason, and since we're just recreating the planar case in the sphere (adjusting the parameters of the model as to localize the 
solutions), we are not in conditions to contribute much to the above discussion.
Instead, we present the static configurations obtained up to charge $B=9$ (Fig. \ref{B>4}) and list the corresponding energies in Table 
\ref{table2}, reinforcing once again the similitudes between the vector meson stabilization of the sigma model and
the more traditional baby-skyrme model.  

\squeezetable
\begin{table}%[h]
\caption{ Energies of localized multi-solitons in the V-M theory. ($\alpha=4.24$ ; $\beta = 1.03$).}
\centering {
\begin{tabular}{c | c | c | l}         \hline \hline
Charge & Energy   & Energy per soliton  & Description\\ 
B      &  $E$     & $E/ B$  & \\ \hline
1      &  18.4    & 18.44   & individual\\
2      &  35.9    & 17.96   & ring \\
3      &  54.2    & 18.05   & 3-soliton \\ 
4      &  71.9    & 17.96   & 2 rings\\
5      &  90.0    & 18.00   & open chain \\
6      &  107.8   & 17.97   & 3 rings\\
7      &  125.8   & 17.97   & closed chain \\
8      &  143.8   & 17.97   & 4 rings  \\
9      &  161.8   & 17.98   & ring + closed chain\\ \hline \hline
%10     &  179.9   & 17.99   & 5 rings \\  \hline \hline

\end{tabular}
}
\label{table2}
\end{table}

%\    \ \vfill

\subsubsection{Large $\beta$ regime}

\begin{figure}
\centering{
\begin{minipage}{4.2cm}
 \subfigure[$\text{  } B=3 \text{ } (D_{2d})$]{\includegraphics[scale=0.2]{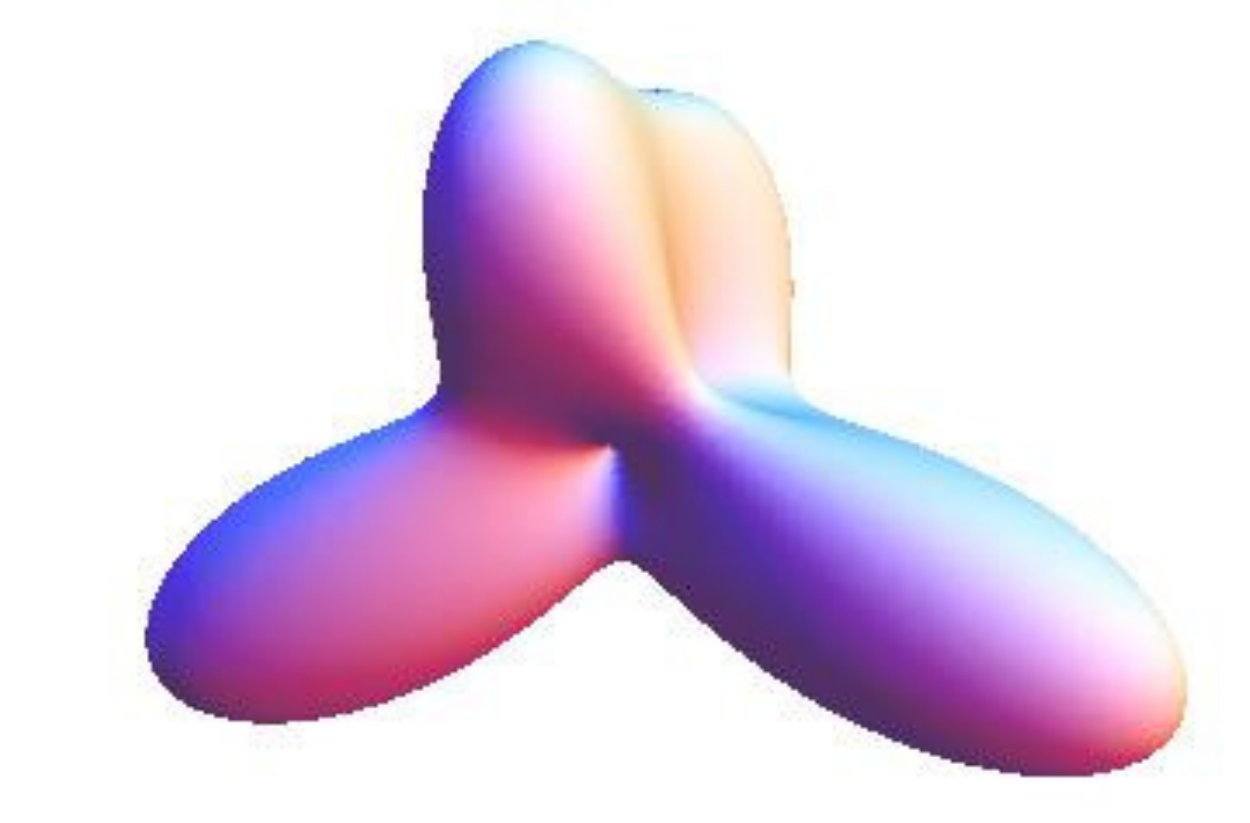}}
\end{minipage}
\begin{minipage}{4.2cm}
 \subfigure[$\text{  } B=4 \text{ } (D_{2d})$]{\includegraphics[scale=0.23]{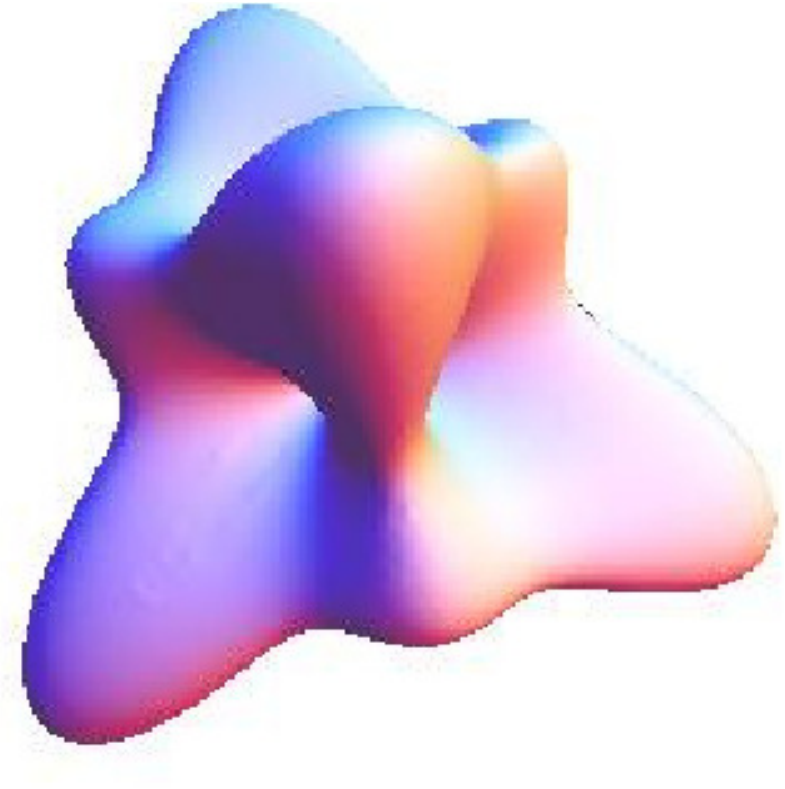}}
\end{minipage}}
\    \ \vfill
\centering{
\begin{minipage}{4.2cm}
\subfigure[$\text{  } B=5 \text{ } (D_{2d})$]{\includegraphics[scale=0.23]{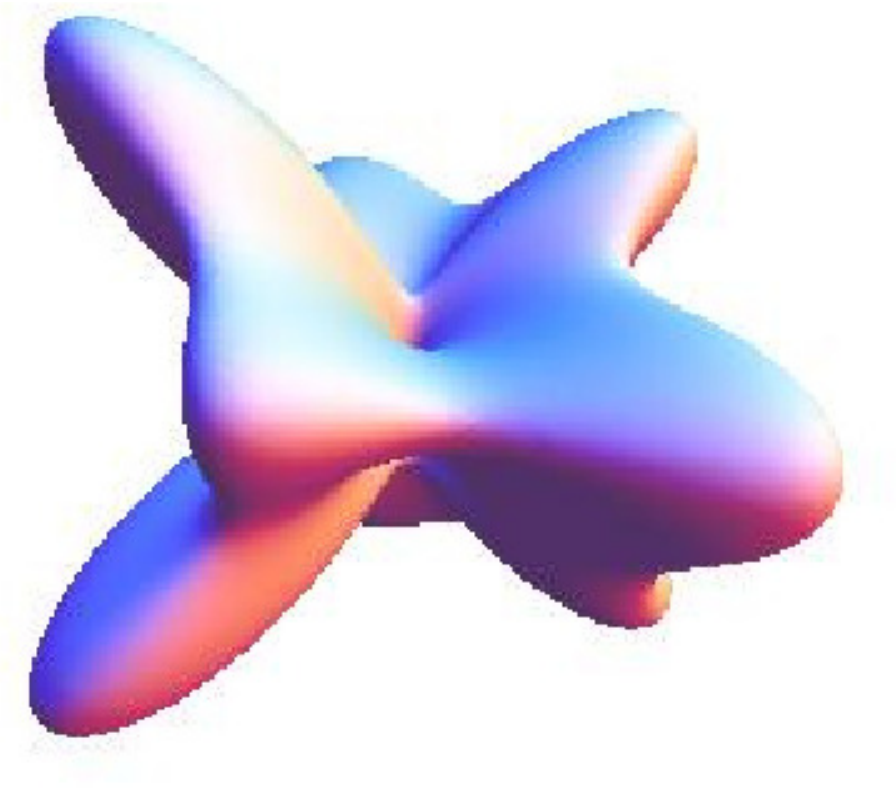}}
\end{minipage}
\begin{minipage}{4.2cm}
\subfigure[$\text{  } B=6 \text{ } (D_{2d})$]{\includegraphics[scale=0.2]{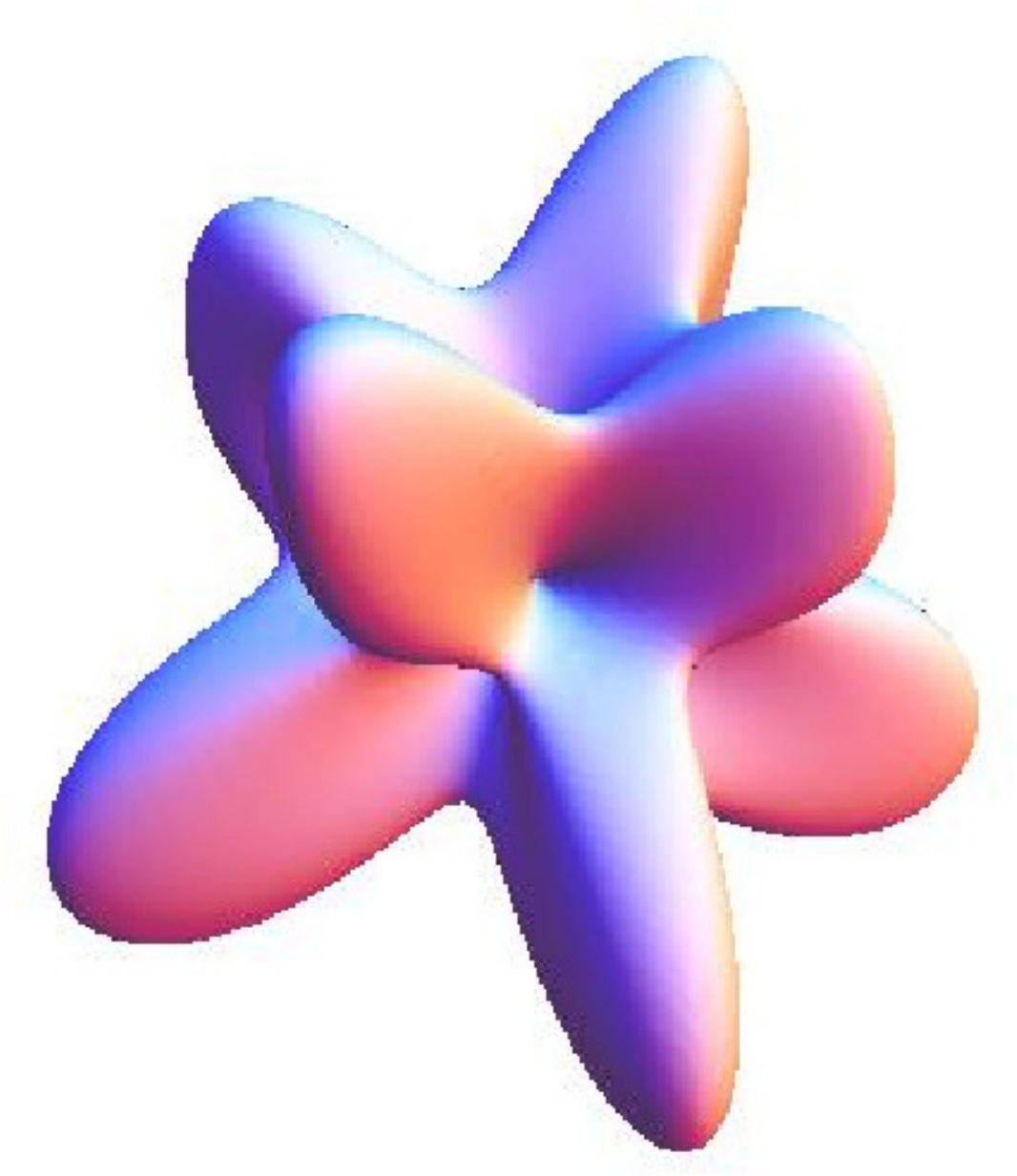}}
\end{minipage}}
\    \ \vfill
\centering{
\begin{minipage}{4.2cm}
\subfigure[$\text{  } B=7 \text{ } (D_{5d})$]{\includegraphics[scale=0.23]{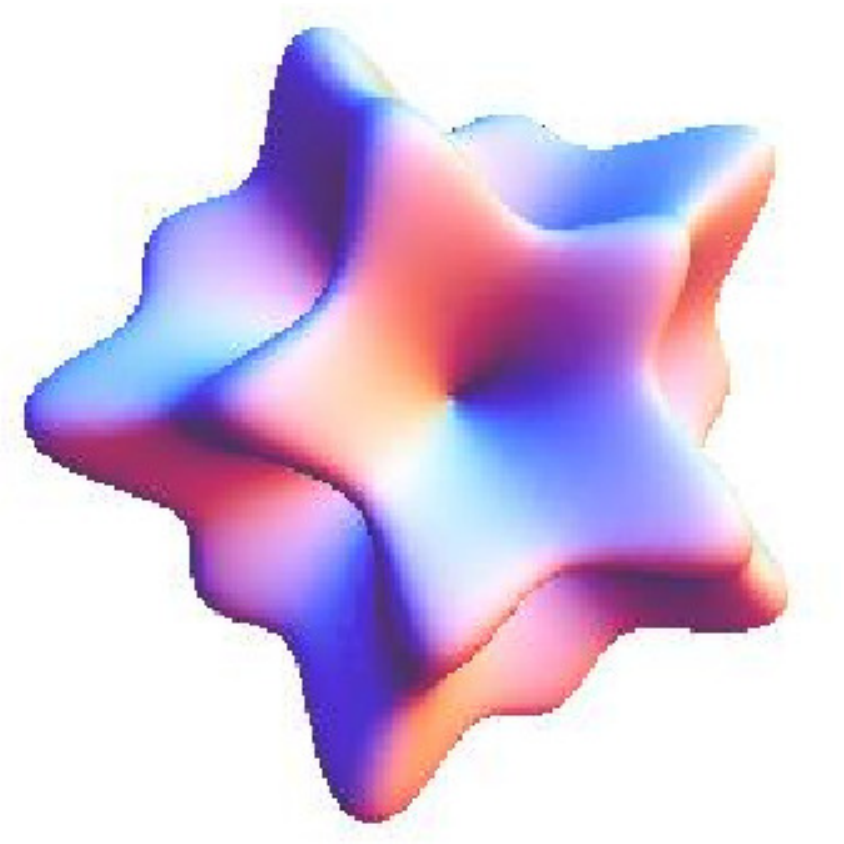}}
\end{minipage}
\begin{minipage}{4.2cm}
\subfigure[$\text{  } B=8 \text{ } (D_{2d})$]{\includegraphics[scale=0.2]{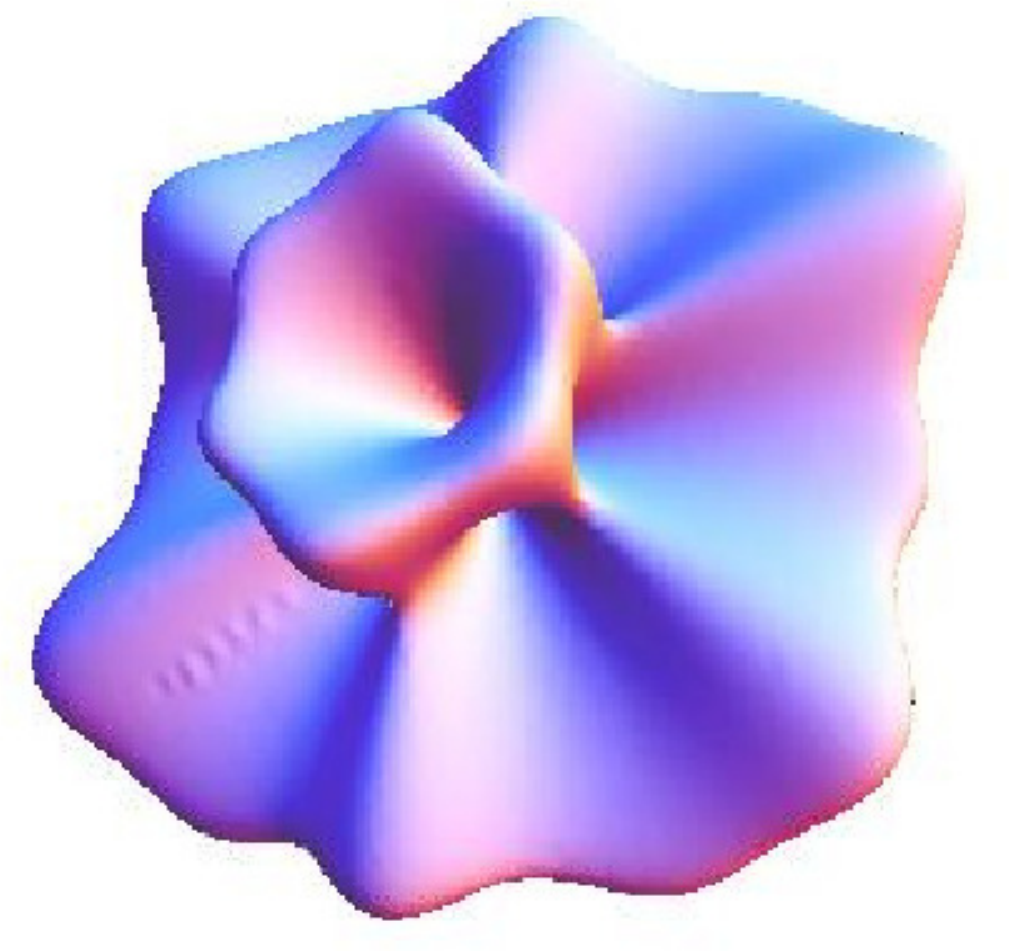}}
\end{minipage}}

 \caption{ High density multi-solitons for charges $3\leq B \leq 8$, with their symmetry groups. 
($\alpha=1.8$, $\beta = 3$ ; $\kappa^2 \simeq 0.1$, $m^2= 0.2$).}
 \label{B>3}
\end{figure}

For higher densities, starting from $\beta$ values close to 3, we find structured solutions like the ones shown on Fig. \ref{B>2}.
This shouldn't be too surprising, since the $V=0$ case represents a limit in which $\beta$ goes to infinity.\\
For the smallest values of $\beta$ within this regime we find multi-solitonic solutions like the ones we display in figure \ref{B>3} 
for the sectors $3 \leq B \leq 11$. They are distorted versions of the topological density plots of Fig. \ref{B>2}.\\
It's important to recall here, that the potential term included breaks the O(3) symmetry of the solutions into the O(2) subgroup. 
So it would be expected for the configurations, now, to have symmetries which are discrete subgroups of O(2) rather than O(3).
This is clearly seen on the solutions of the sectors $B=3$, $B=4$ and $B=7$, having platonic symmetries  when $V=0$ (Tetrahedral, Cubic and 
Icosahedral, respectively), and now exhibiting dihedral symmetries with a preferred axis.

For larger values of $\beta$, these effects become weaker and the multi-solitonic solutions approach those of section $V=0$ (Fig. \ref{B>2}).
 
%\begin{table}%[h]
%\caption{ Distorted Symmetries. ($\alpha=1.8$, $\beta = 3$ ; $\kappa \simeq 0.1$, $m^2= 0.2$).}
%\centering {
%\begin{tabular}{|c|c|c|c|c|c|c|}         \hline 
%                                      & $B=3$    & $B=4$    & $B=5$    & $B=6$    & $B=7$    & $B=8$    \\ \hline
%$V=0$ (masslees pions)                & $O_h$    & $O_h$    & $D_{4d}$ & $D_{4d}$ & $I_h$    & $D_{6d}$ \\ \hline
%$V\neq0$ ($\alpha=1.8$ ; $\beta = 3$) & $D_{2d}$ & $D_{2d}$ & $D_{2d}$ & $D_{2d}$ & $D_{5d}$ & $D_{2d}$ \\ \hline 
%
%\end{tabular}}
%\label{Symmetries}
%\end{table}

\section{Summary and Conclusions}

We have studied the vector meson stabilization of the sigma model on the two-sphere, performing a numerical implementation of the problem
which allows us to find static multi-soliton solutions and to check dynamically whether such solutions were stable.  

We were able to compare our solutions up to charge $B=14$, for the massless pion case ($V=0$), with the ones obtained in the 
Baby-Skyrme model, finding an incredibly correspondence between the two models, not only in a qualitative ground (symmetries)
but also on quantitative level (energies).
These solutions were found to have generally the same symmetries as corresponding multi-skyrmions of the 3D Skyrme model, and it was 
suggested in the literature a strong connection between them. 
We didn't pursue these arguments further to explore the possible effects that a non-zero pion mass would have, because we believe this
association between the two-dimensional version on the sphere and the three-dimensional model is only true for maps that can be well approximated
by rational maps (which was the case here), but the introduction of the potential term will modify this situation and a more careful analysis
would be then required. 
We defer such analysis to a forthcoming work, in which we shall be interested on the possibility of extending the studies \cite{ mass, pions} 
for the theory with vector mesons and no Skyrme term. In those studies, it is shown that there is an important qualitative 
difference between multi-skyrmions (in the standard Skyrme model) with massive or massless pions. For sufficiently large pion mass or baryon number,
the structures collapse to form qualitatively different stable Skyrmion solutions.

We have also explored here, the inclusion of a potential term (the more traditional one being that of a pion mass) and 
the interplay generated between the new length scale it introduce and the natural spatial scale of the two-sphere. 
We have identified the relevant parameter associated with the concept of a soliton density on space, and we have basically observed two regimes in which
very different types of solutions are found. Although, the transition is not sharply defined.\\
The first regime, is a low density phase where we have found localized solutions that allowed us to recreate the planar two-dimensional case on the sphere,
and hence, allowing us to compare our solutions with the ones present in the literature (both for baby-skyrmion and vector meson stabilization),
finding good qualitative agreement and thus reinforcing the great similarities between the two models.\\
While the second regime of higher soliton densities, shows structured solutions covering the entire sphere that are very similar to those encountered for 
the massless ($V=0$) case. They seem to be distorted and to have loose their symmetry when close to the phase transition ($\beta \sim 3$).

It is worth mentioning, that we already have here the numerical infrastructure to study the dynamical aspects of the model as well, in particular
the scattering and annhilitation process like the ones presented in refs. \cite{ Amado-1, Amado-2}, where it has been found a rich variety of phenomena
and an intimate connection with the Skyrme model.

\bibliographystyle{unsrt} %plain
\bibliography{biblio}

\end{document}